\documentclass[
aps,
reprint,
superscriptaddress,
longbibliography,
dblfloatfix,
]{revtex4-2}

\usepackage[T1]{fontenc}
\usepackage[utf8]{inputenc}
\usepackage{lmodern}
\usepackage{microtype}
\usepackage{graphicx}
\usepackage{bm}
\usepackage{amsmath,amssymb,mathtools}
\usepackage{siunitx}
\usepackage{booktabs}
\usepackage{dcolumn}
\usepackage{array,multirow}
\usepackage{hyperref}
\usepackage[capitalise]{cleveref}
\usepackage[acronym]{glossaries}
\usepackage{tabularx}

\makeatletter
\def\place@bibnumber#1{#1.}
\makeatother
\setcitestyle{super,comma}

\makeglossaries

\newacronym{laspm}{LA-SPM}{light-assisted scanning-probe microscopy}
\newacronym{pdh}{PDH}{Pound-Drever-Hall}
\newacronym{cqed}{cQED}{cavity quantum electrodynamics}
\newacronym{com}{cOM}{cavity optomechanics}
\newacronym{dbr}{DBR}{distributed Bragg reflector}
\newacronym{fir}{FIR}{finite impulse response}
\newacronym{fpga}{FPGA}{field-programmable gate array}
\newacronym{eom}{EOM}{electro-optic modulator}
\newacronym{fwhm}{FWHM}{full width at half maximum}
\newacronym{sof}{SOF}{side-of-fringe}

\newcommand{\capshort}[1]{\glsentryshort{#1}}

\newcommand{\singlecolwidth}{85mm}
\newcommand{\doublecolwidth}{170mm}
\DeclareMathSymbol{\mathhyphen}{\mathord}{operators}{`-}

\makeatletter
\def\blfootnote{\gdef\@thefnmark{}\@footnotetext}

\long\def\@makefntext#1{%
 \parindent 0pt%
 \noindent
 \hb@xt@0pt{\hss}% no marker box
 #1%
}
\makeatother

\newlength{\capheight}
\newcommand{\topalignmark}[1]{%
 \begingroup
 \sbox0{X}% reference cap-height box
 \sbox2{#1}% symbol box
 \raisebox{\dimexpr\ht0-\ht2\relax}{\usebox2}%
 \endgroup
}

\newlength{\blfnlabelwidth}
\newcommand{\blfnotetext}[2]{%
 \blfootnote{%
  \hspace{-1.1em}
  \noindent
  \makebox[\blfnlabelwidth][r]{\topalignmark{#1}}%
  \hspace{0.5em}%
  \parbox[t]{\dimexpr\linewidth-\blfnlabelwidth-0.5em\relax}{#2}%
 }%
}

\begin{document}

\title{A dry cryogenic platform with picometre stability for nanoscale and quantum technologies}

\begin{abstract}

\begin{center}
\begin{minipage}{0.75\textwidth}
\centering
{\fontsize{10.5}{11}\selectfont
\vspace{0.5em}
William G. Roth,\textsuperscript{1,\hyperlink{fn:equal}{*}}
Andreas Thurn,\textsuperscript{1,\hyperlink{fn:equal}{*},\hyperlink{fn:corr}{\textdagger}}
Alexander M. Stramma,\textsuperscript{1, \hyperlink{fn:paddr}{\textdaggerdbl}}
Amit R. Dhawan,\textsuperscript{2}
Jason M. Smith,\textsuperscript{2}
Dorian A. Gangloff,\textsuperscript{1,\hyperlink{fn:corr}{\textdagger}} and
Mete Atat\"ure\textsuperscript{1,\hyperlink{fn:corr}{\textdagger}}
}
\end{minipage}

{\vspace{0.6em}
\fontsize{9.5}{10}\selectfont\itshape{
$^{1}$Cavendish Laboratory, University of Cambridge, J. J. Thomson Avenue, Cambridge CB3 0US, UK\\
$^{2}$Department of Materials, University of Oxford, Parks Road, Oxford OX1 3PH, UK
}}
\end{center}
\vspace{-0.2em}

Mechanically ultrastable environments within maintenance-free dry cryostats are needed for high-precision surface probe microscopy experiments, as well as next-generation quantum technologies. The most demanding applications require picometre-scale precision combined with multi-axis and millimetre-scale scanning range, while offering access to optical, microwave, and radiofrequency control and a vector magnetic field. We demonstrate a versatile platform for such systems based on passive and active vibration control for a commercial dry cryostat. Using a tunable optical microcavity as a sensor, we report vibration levels of \SI{4.2}{pm} root mean square at \SI{2.3}{K} over one hour without postselection. We further demonstrate drift-free long-term operation with stability below \SI{8.6}{pm} RMS over 12 hours. Our system enables continuous and automated operation for next-generation high-throughput applications without the need for liquid helium infrastructure. Requiring only standard machining facilities and commercially available hardware, the design can be readily adapted for cryogenic optical, electrical and magnetic high-precision measurements across quantum science, surface science, and nanotechnology.

\end{abstract}

\maketitle

%%==================================%%
%% Main %%
%%==================================%%
\label{sec1}

\begingroup
\setlength{\footnotesep}{0.6\baselineskip}
\blfnotetext{\hypertarget{fn:equal}{\scalebox{0.7}{*}}} {These authors contributed equally to this work.}
\blfnotetext{\hypertarget{fn:corr}{\scalebox{0.7}{\textdagger}}}{Corresponding authors: at2116@cam.ac.uk; dag50@cam.ac.uk; ma424@cam.ac.uk.}
\blfnotetext{\hypertarget{fn:paddr}{\scalebox{0.7}{\textdaggerdbl}}}{\kern-0.25\baselineskip Present address: QuTech and Kavli Institute of Nanoscience, Delft University of Technology, Delft, The Netherlands.}
\endgroup

Experimental platforms at the forefront of nanoscale and quantum technologies face increasingly stringent, simultaneous requirements. Cryogenic temperatures at or below liquid helium temperature are often necessary, for example, to preserve the coherence of quantum systems and to reveal novel behaviour of materials~\cite{takahashiQuenchingSpinDecoherence2008,pingaultCoherentControlSiliconvacancy2017,limCoherentSpinDynamics2018,welinskiElectronSpinCoherence2019,liHyperfineStructureCoherent2020,merkelCoherentPurcellEnhancedEmission2020,guoMicrowaveBasedQuantumControl2023a,harrisCoherenceGroupIVColor2024,zhangOpticalSpinCoherence2024,tiranovSubsecondSpinLifetimelimited2026}. Compatibility with a magnetic field is also essential for studying magneto-optical phenomena~\cite{dapolitoInfraredNanoimagingDirac2023} and, in combination with electrical and optical field delivery, for multi-modal nanoscale probing and quantum control of qubit systems~\cite{guoMicrowaveBasedQuantumControl2023a, pressCompleteQuantumControl2008, rancicCoherenceTimeSecond2018}. 
Next-generation platforms must provide these capabilities alongside multiple degrees of freedom for mechanical alignment~\cite{birkbeckQuantumTwistingMicroscopy2025}. Such in situ tunability is indispensable for scanning-probe architectures~\cite{shiotariPicocavityEnhancedRamanSpectroscopy2025, wittnevenLowtemperatureUltrahighvacuumScanning1997, shiotariScatteringNearfieldOptical2025,kasaiDevelopmentNear5KelvinCryogenfree2022, chaudharyLowNoiseCryogenfree2021, mengAtomicallyResolvedProbetype2019, huangCryogenFreeSpin2022,birkbeckQuantumTwistingMicroscopy2025} and for efficient light-matter interfaces such as tunable optical microcavities~\cite{bogdanovicDesignLowtemperatureCharacterization2017, fontanaMechanicallyStableTunable2021,rufResonantExcitationPurcell2021,vadiaOpenCavityClosedCycleCryostat2021,pallmannHighlyStableFully2023,fisicaroActiveStabilizationOpenaccess2024,herrmannLowtemperatureTunableMicrocavity2024,herrmannCoherentCouplingDiamond2024,ulanowskiSpectralMultiplexingRareEarth2024,fischerSpinPhotonCorrelationsPurcellenhanced2025}, where quantum emitters need to be positioned relative to the optical cavity mode. 
However, this mechanical tunability is in direct conflict with the need for picometre-level mechanical stability, which is necessary to maintain the resonance condition in high-finesse microcavities~\cite{janitzCavityQuantumElectrodynamics2020a,pallmannHighlyStableFully2023} and to control the tip–sample distance with angstrom-level precision in advanced scanning-probe systems~\cite{shiotariPicocavityEnhancedRamanSpectroscopy2025,decamposferreiraResonantTipEnhancedRaman2024,leeVisualizingVibrationalNormal2019,zhangChemicalMappingSingle2013, yangSubnanometreResolutionSinglemolecule2020, imadaSinglemoleculeLaserNanospectroscopy2021}.
Beyond these mechanical considerations, free-space optical access is often required to achieve efficient mode matching~\cite{gallegoHighfinesseFiberFabry2016, yurgensCavityassistedResonanceFluorescence2024}, to avoid the phase noise introduced by optical fibres~\cite{yurgensCavityassistedResonanceFluorescence2024, fischerSpinPhotonCorrelationsPurcellenhanced2025} and to support multiple optical beam paths~\cite{hillenbrandVisibletoTHzNearfieldNanoscopy2025}.

Enabling such systems in dry cryostats is highly desirable, as they provide continuous and automated operation for next-generation high-throughput applications. These cryostats, however, rely on cryocoolers that generate spectrally broadband mechanical vibrations~\cite{vadiaOpenCavityClosedCycleCryostat2021,chijiokeVibrationSpectrumPulsetube2010}. Without effective suppression, this noise precludes experiments requiring sub-nanometre stability. Consequently, considerable effort is being directed towards developing ultrastable and versatile platforms based on dry cryostats that provide mechanical stability on par with liquid helium bath systems~\cite{najerGatedQuantumDot2019,greuterSmallModeVolume2014, shiotariPicocavityEnhancedRamanSpectroscopy2025, wittnevenLowtemperatureUltrahighvacuumScanning1997, shiotariScatteringNearfieldOptical2025}.

\begin{figure*}
\includegraphics[width=\textwidth, height=0.49\textheight, keepaspectratio]{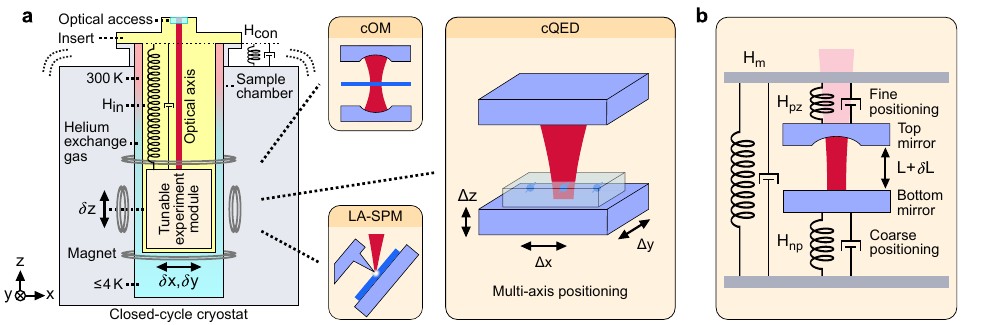}
\caption{\textbf{The challenge of achieving picometre-scale surface stability in vibrating closed-cycle cryostats.} 
\textbf{a}, Sketch of a closed-cycle cryostat with a removable insert (shaded in yellow) that provides optical access (red line) to a tunable experiment module. The cryostat creates a vertical temperature gradient (colour
gradient from blue to red) along the sample chamber wall, which is thermally coupled to the insert and the experiment module via a helium exchange gas. The mechanical connections from the upper end of the insert to the cryostat and to the experiment module are described by transfer functions $H_{\rm con}$ and $H_{\rm in}$, respectively. Vibrations from the cryostat's cryocooler introduce translational ($\delta x$, $\delta y$, $\delta z$) mechanical noise at the experiment module. Together, $H_{\rm con}$ and $H_{\rm in}$ attenuate this noise and enable tunable experiments for a wide range of applications that require picometre-level stability, including cavity quantum electrodynamics (\capshort{cqed}), cavity optomechanics (\capshort{com}), and light-assisted scanning probe microscopy (\capshort{laspm}).
\textbf{b}, Schematic of the three-axis tunable microcavity serving as a high-precision vibration sensor along the $z$-axis. Coarse positioning of the bottom mirror ($\Delta x$, $\Delta y$, $\Delta z$) is provided by a nanopositioner stack, while fine-tuning of the cavity length is enabled by a piezoelectric actuator connected to the top mirror. Vibrations transmitted to the experiment module result in detunings ($\delta L$) from the cavity resonance length ($L$). These detunings are governed by the transfer functions of the nanopositioner stack ($H_{\text{np}}$), the piezoelectric actuator ($H_{\text{pz}}$) and the connecting components ($H_{\text{m}}$).}
\label{fig:1}
\end{figure*}

Recent advances in cryogenic scanning probe microscopy platforms have satisfied some of the above-stated requirements. For instance, cryogenic scanning tunnelling microscopy setups have achieved picometre-level stability in closed-cycle environments~\cite{esserUltrahighVacuumScanning2024,maDevelopmentCryogenfreeSub32023, hackleyHighstabilityCryogenicScanning2014}, but these designs are often bulky and incompatible with optical access~\cite{kasaiDevelopmentNear5KelvinCryogenfree2022, chaudharyLowNoiseCryogenfree2021, mengAtomicallyResolvedProbetype2019, huangCryogenFreeSpin2022}. First demonstrations of \gls{laspm} setups with free-space optical access in closed-cycle cryostats~\cite{dapolitoInfraredNanoimagingDirac2023,shiotariScatteringNearfieldOptical2025}, on the other hand, have so far been limited to stability levels on the order of several nanometres. This remains insufficient for more demanding methods such as tip-enhanced Raman scattering (TERS)~\cite{shiotariPicocavityEnhancedRamanSpectroscopy2025,decamposferreiraResonantTipEnhancedRaman2024,leeVisualizingVibrationalNormal2019,zhangChemicalMappingSingle2013} and tip-enhanced photoluminescence (TEPL)~\cite{yangSubnanometreResolutionSinglemolecule2020, imadaSinglemoleculeLaserNanospectroscopy2021} for sub-molecular-resolution spectroscopy~\cite{yangSubnanometreResolutionSinglemolecule2020, leeVisualizingVibrationalNormal2019, shiotariPicocavityEnhancedRamanSpectroscopy2025, decamposferreiraResonantTipEnhancedRaman2024, zhangChemicalMappingSingle2013}. Such techniques require mechanical stability on the order of tens of picometres in all three spatial directions.

Another highly active area of research focuses on developing tunable microcavity platforms in closed-cycle cryostats for \gls{cqed} experiments. The high finesse and resulting narrow linewidths of these microcavities impose strict stability constraints on the cavity length, which are on the order of \SI{10}{pm} RMS~\cite{janitzCavityQuantumElectrodynamics2020a,pallmannHighlyStableFully2023}. Transverse to the optical axis, the stability requirements are typically less severe and are strongly dependent on the specific application. These spectrally and spatially tunable systems can be coupled to a wide range of solid-state emitters~\cite{wangCoherentCouplingSingle2017,merkelCoherentPurcellEnhancedEmission2020,casaboneDynamicControlPurcell2021}, including quantum dots~\cite{najerGatedQuantumDot2019,tommBrightFastSource2021,antoniadisCavityenhancedSingleshotReadout2023} and diamond colour centres~\cite{herrmannCoherentCouplingDiamond2024, yurgensCavityassistedResonanceFluorescence2024}. Such cavity-coupled emitters hold great promise for photonic quantum computing~\cite{knillSchemeEfficientQuantum2001,kokLinearOpticalQuantum2007} and quantum networks~\cite{janitzCavityQuantumElectrodynamics2020a,rufQuantumNetworksBased2021a}. Further, due to their small mode volume, tunable high-finesse microcavities also enhance the optomechanical coupling to low-mass mechanical resonators~\cite{flowers-jacobsFibercavitybasedOptomechanicalDevice2012,pfeiferAchievementsPerspectivesOptical2022,ruelleTunableFiberFabry2022}. This makes them attractive platforms for \gls{com}~\cite{aspelmeyerCavityOptomechanics2014,ruelleTunableFiberFabry2022} and for integration with tunable optomechanical scanning-probe systems for nanoscale precision sensing~\cite{heOptomechanicalAtomicForce2021,foglianoMappingCavityOptomechanical2021,gislerEnhancingMembranebasedScanning2024}.

Among tunable open-access optical microcavities with at least three spatial tuning axes, the best reported closed-cycle cavity-length stability is \SI{10.6}{pm} RMS with active stabilisation over a full cryocooler cycle at \SI{4}{K}~\cite{fisicaroActiveStabilizationOpenaccess2024}. However, this implementation is not compatible with applied magnetic fields, and the same limitation applies to other reported designs~\cite{fontanaMechanicallyStableTunable2021, vadiaOpenCavityClosedCycleCryostat2021}. Other approaches have simultaneously achieved magnetic-field compatibility and $25\,\mathrm{pm}$ RMS passive stability, but at the cost of reduced thermal anchoring, resulting in an elevated base temperature of \SI{8}{K}~\cite{herrmannLowtemperatureTunableMicrocavity2024,herrmannCoherentCouplingDiamond2024}. In contrast, limiting motion to a single degree of freedom with a highly constrained movement range of approximately \SI{20}{nm} has enabled a stability of \SI{1.5}{pm} RMS at \SI{1.7}{K}~\cite{merkelCoherentPurcellEnhancedEmission2020,merkelEnhancingEmissionCoherence2021}, but this trade-off precludes experiments requiring multi-axis alignment, as is needed, for example, for sensing and for optimal positioning of single quantum emitters with respect to the cavity mode. To our knowledge, no prior system~\cite{bogdanovicDesignLowtemperatureCharacterization2017, fontanaMechanicallyStableTunable2021,rufResonantExcitationPurcell2021,vadiaOpenCavityClosedCycleCryostat2021,pallmannHighlyStableFully2023,fisicaroActiveStabilizationOpenaccess2024,herrmannLowtemperatureTunableMicrocavity2024,herrmannCoherentCouplingDiamond2024,merkelCoherentPurcellEnhancedEmission2020,ulanowskiSpectralMultiplexingRareEarth2024,fischerSpinPhotonCorrelationsPurcellenhanced2025,kosterHighcooperativityCouplingSpinresolved2026} meets all requirements of combining operation in a closed-cycle cryostat at $\leq\SI{4}{K}$, free-space optical access, magnetic-field compatibility, three-axis tunability, and $<\SI{10}{pm}$ cavity-length stability.

Here, we introduce a versatile vibration-isolation architecture that meets these requirements. Using a three-axis tunable, open-access optical microcavity as a demonstrator, we achieve a passive cavity-length stability of \SI{13.9}{pm} RMS at \SI{4}{K} measured over $1\,\text{hour}$ within a frequency range from $\SI{280}{\micro Hz}$ to $\SI{50}{kHz}$ without postselection. Active feedback improves this stability to \SI{8.1}{pm} RMS and enables drift-free, continuous operation, as demonstrated over $12\,\text{hours}$. With an upgraded implementation, we further reduce this figure to \SI{4.2}{pm} RMS at a temperature of \SI{2.3}{K} with an applied vector magnetic field of \SI{283}{mT}. This corresponds to an improvement in cavity-length stability by three orders of magnitude relative to an otherwise identical microcavity in the same cryostat without our vibration-isolation architecture. We note that the residual vibration levels are on par with liquid helium bath cryostats~\cite{yurgensCavityassistedResonanceFluorescence2024}.

%%%%%%%%%%%%%
% Subsection 1
%%%%%%%%%%%%%
\section*{Results}
\subsection*{The mechanical challenge}
\label{sec:label1}

\begin{figure*}[t!]
\includegraphics[width=\textwidth, height=0.48\textheight, keepaspectratio]{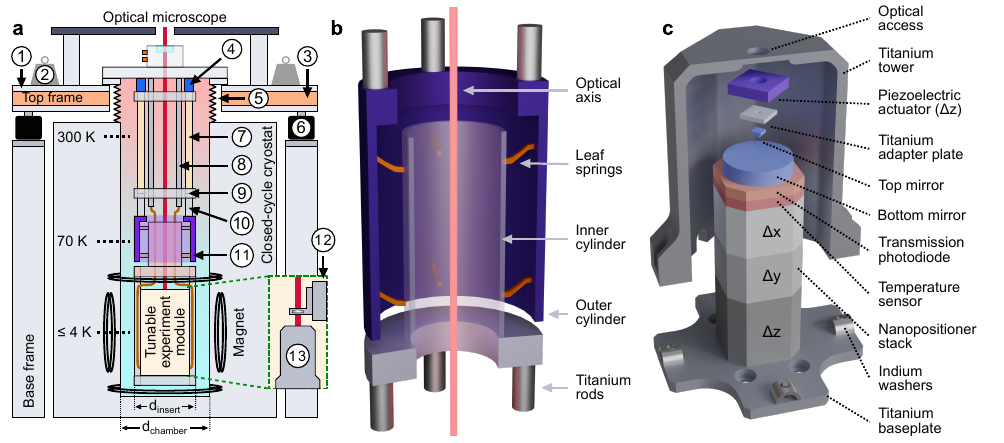}
\caption{\textbf{Mechanical design of a dry cryogenic platform with picometre-scale stability and multiple degrees of freedom.}
\textbf{a}, Schematic illustration of the vibration isolation architecture. A custom insert of diameter $d_{\mathrm{insert}}$ holds the tunable experiment module within a sample chamber of diameter $d_{\mathrm{chamber}}$, positioning it at the magnetic-field centre of a superconducting vector magnet. An aluminium frame (1) mass-loaded with lead weights (2) and filled with silicone (3) rests on air isolators (6) and supports both the insert and an optical microscope. Edge-welded bellows (5) provide a compliant mechanical connection between the insert and the cryostat. Rubber isolators (4) decouple the cageplates (9), as well as the fibreglass (7) and titanium rods (10) from the supporting frame. A custom spring interconnect (11) provides further isolation and, in addition, decouples the experiment from the semi-rigid microwave lines (8). In this work, the tunable experiment module is an optical microcavity (13), as shown by the inset (green dashed lines). Mode matching to external optics is provided by an aspheric lens mounted to a nanopositioner (12). \textbf{b}, Sketch of the custom spring interconnect (11), shown in \textbf{a}. Two concentric cylinders are connected to each other via two axially separated sets of four leaf springs. \textbf{c}, Exploded view of the three-axis tunable microcavity. The bottom mirror can be positioned by a nanopositioner stack with millimetre-scale range in all three directions ($\Delta x$, $\Delta y$, $\Delta z$). For picometre-precise adjustments of the cavity length, a piezoelectric actuator is used to move the top mirror along the optical axis ($\Delta z$). Plastically deformable indium washers enable angular alignment of the top mirror with respect to the bottom mirror.} 
\label{fig:2}
\end{figure*}

Figure~\ref{fig:1}a illustrates the central challenge of integrating a tunable experiment module that demands picometre-scale stability and free-space optical access into a closed-cycle cryostat. In our demonstration, we use an Attocube 2100 cryostat, but we note that our design can be straightforwardly adapted to a range of other cryostats. The insert (shaded in yellow) is a removable assembly that supports the experiment module and is mounted to the cryostat via a vacuum-compatible mechanical connection. The connections from the upper end of the insert to the cryostat and to the experiment module are described by the transfer functions $H_{\rm con}$ and $H_{\rm in}$, respectively, schematically represented in Fig.~\ref{fig:1}a by damped springs. Together, these connections ($H_{\rm con}$ and $H_{\rm in}$) form the mechanical interface between the cryostat and the experiment module. 

Using a closed-cycle system based on a pulse-tube cryocooler, the cryostat cools a superconducting vector magnet to \SI{4}{K}. In parallel, the bottom section of the sample chamber can be temperature-controlled between \SI{1.7}{K} and \SI{300}{K}, while its top section remains at \SI{300}{K}. We set the bottom section of the sample chamber to temperatures of \SI{4}{K} and below, which results in a temperature gradient along the sample chamber, as indicated by the vertical colour gradient from blue to red in Fig.~\ref{fig:1}a. The insert and the experiment module are thermally coupled to the sample chamber via a helium exchange gas. This creates a temperature gradient along the insert and cools the experiment module to the selected cryogenic operating temperature.

In the pulse-tube cryocooler, periodic compression and expansion of helium gas at a frequency of \SI{1.4}{Hz} produce pulses of mechanical vibrations that are spectrally broadband, with frequency components extending to a few kHz~\cite{vadiaOpenCavityClosedCycleCryostat2021, chijiokeVibrationSpectrumPulsetube2010} (see Supplementary Note~1). These vibrations couple to the rest of the cryostat, and, from there, they reach the experiment module through the mechanical interface, but not through the helium exchange gas. At the experiment module, these vibrations manifest as translational ($\delta x$, $\delta y$, $\delta z$) mechanical noise, as shown in Fig.~\ref{fig:1}a. This noise drives relative motion between components within the module and can also misalign the module as a whole with respect to the optical axis. 

The tunable experiment module can be adapted for a range of configurations, from \gls{com} and \gls{laspm} experiments to \gls{cqed} with solid-state emitters, as illustrated in Fig.~\ref{fig:1}a. These share a common demand on the mechanical interface: it must enable relative component positioning stable to within tens of picometres along the axes to which the measured signal is most sensitive, retain tunability along these and additional degrees of freedom, and preserve the module's alignment to the optical axis. Here, we address the case of a single highly sensitive axis, as encountered in \gls{cqed} with solid-state emitters.

We implement a tunable optical microcavity with a plano-concave geometry, as shown in Fig.~\ref{fig:1}b. In our design, the bottom mirror is mounted on a stack of translation stages, referred to as nanopositioner stack, for coarse positioning along the $x$, $y$, and $z$ directions, while a piezoelectric actuator connected to the top mirror provides fine-tuning of the cavity length with picometre resolution. This microcavity design serves a dual role. In this work, it acts as a high-precision vibration sensor along the $z$-axis, enabling us to quantify platform stability and to actively stabilise the mirror separation. In the future, placing a substrate hosting solid-state emitters on the bottom mirror would allow individual emitters to be aligned spatially and spectrally with the cavity mode for optimal coupling, making this design a versatile architecture for \gls{cqed}.

 The nanopositioner stack and the piezoelectric actuator are characterised by mechanical transfer functions $H_{\rm np}$ and $H_{\rm pz}$, respectively. Together, $H_{\rm np}$, $H_{\rm pz}$, and the transfer function of the mechanical components connecting them ($H_{\rm m}$) determine how vibrations transmitted to the experiment module result in displacement noise $\delta L$ in the cavity length $L$. 

Since the mechanical resonances of typical nanopositioner stacks lie within the dominant noise band (up to \SI{3}{\kilo\hertz}, see Supplementary Note~1) of pulse-tube cryocoolers~\cite{vadiaOpenCavityClosedCycleCryostat2021, chijiokeVibrationSpectrumPulsetube2010}, these resonances contribute strongly to $\delta L$. To avoid introducing additional resonances within this band, the connecting components ($H_{\rm m}$) thus need to be sufficiently stiff, with mechanical resonance frequencies ideally well above $\SI{3}{\kilo\hertz}$. 

%%%%%%%%%%%%%
% Subsection 2
%%%%%%%%%%%%%
\subsection*{The mechanical solution}
\label{sec:label2}

Figure~\ref{fig:2}a shows a schematic of our vibration-isolation architecture. A custom insert places the tunable experiment module in the magnetic field centre of a superconducting vector magnet. The experiment module is thermalised with the walls of the lower section of the sample chamber, which is held at a temperature of $\leq$~\SI{4}{K}, via the helium exchange gas. In this work, the module is a tunable open microcavity, as illustrated by the inset (green dashed lines) and in Fig.~\ref{fig:1}b. Mode matching to external optics is provided by an aspheric lens with a focal length of \SI{7.5}{mm} and a numerical aperture of 0.3, mounted to a vertically oriented low-temperature nanopositioner with a \SI{5}{mm} movement range.

 The insert has a diameter of $d_{\mathrm{insert}}=\SI{41.4}{mm}$ and thus provides $\SI{4}{mm}$ of clearance to the wall of the sample chamber, which has an inner diameter of $d_{\mathrm{chamber}}=\SI{49.7}{mm}$. This large clearance prevents unintended contact with the sample chamber wall. When the insert is correctly aligned with respect to the sample chamber, mechanical coupling between the insert and the cryostat is thus limited to a single compliant element: soft, edge-welded bellows. These bellows are made out of non-magnetic steel and provide a gastight yet flexible mechanical connection between the cryostat and the upper end of the insert. In combination with the mass of the insert and the top frame, these bellows act as a mechanical low-pass filter, which reduces vibration transfer from the cryostat to the insert along all three axes ($x$, $y$ and $z$). Together, these bellows and the insert form the mechanical interface between the experiment module and the cryostat. 

The insert and the optical microscope are supported by a rigid aluminium frame (top frame), which is stable and drift-free. To dampen the frame's mechanical resonances, its originally hollow aluminium profiles are filled with silicone with a Shore hardness of A40. In addition, \SI{25}{kg} of lead weights are used to increase the mass of this support frame and to balance its weight distribution. Passive air dampers with a vertical cut-off frequency of \SI{3.5}{Hz} and a transmissibility at resonance of \SI{24}{dB} decouple the top frame from a rigid ``base frame'', which surrounds the cryostat and supports the weight of the top frame assembly.

Along the insert, vibration isolation is implemented in two stages, leveraging the vertical thermal gradient inside the sample chamber. At the top, near room temperature, the damping properties of elastomers are preserved, allowing the use of rubber connectors with a Shore hardness of A60. These rubber connectors reduce vibration transfer between the insert's fibreglass rods and the top frame along all three axes ($x$, $y$ and $z$). Closer to the experiment module, near \SI{70}{K}, a custom spring interconnect decouples the experiment from residual vibrations along the $z$-axis, transmitted along the fibreglass rods, titanium rods, and the semi-rigid microwave lines.

The interconnect, illustrated in Fig.~\ref{fig:2}b, connects an outer and an inner cylinder via two axially separated groups of four symmetrically arranged leaf springs. This geometry provides mechanical isolation along the $z$-axis, while remaining stiff in all other directions and against rotations, thereby maintaining alignment with the optical axis. This design was guided by the principles outlined in the work by Fontana et al.~\cite{fontanaMechanicallyStableTunable2021}. The leaf springs are made of phosphor bronze, which retains its ductility at cryogenic temperatures~\cite{simonPropertiesCopperCopper1992}, while the cylinders were machined from aluminium. When supporting a \SI{1}{kg} mass, the spring interconnect has a natural frequency~\cite{bauchauEulerBernoulliBeamTheory2009} of \SI{20}{Hz}.

Figure~\ref{fig:2}c illustrates the mechanical design of the open microcavity. The cavity has a length of \SI{5.614(2)}{\micro m} (see Supplementary Note~2). The mirror substrates are made out of fused silica and are coated with a \gls{dbr} layer structure (see Methods). The symmetric coatings yield a measured finesse of $F=3210(25)$ (see Supplementary Note~2) at a wavelength of $\lambda=\SI{633}{nm}$. This corresponds to a spatial linewidth of $\delta L_{\mathrm{FWHM}} = \lambda/(2F)=\SI{98.5(8)}{pm}$, which describes the \gls{fwhm} of the reflection dip when the cavity length is scanned across a cavity resonance at a fixed laser wavelength. The bottom mirror is mounted on a low-temperature nanopositioner stack that provides a movement range of \SI{3}{mm} in all three directions ($\Delta x$, $\Delta y$, and $\Delta z$). A photodiode and a sensor are placed below the bottom mirror to monitor the transmitted light intensity and the local temperature. The top mirror is attached to a titanium adapter plate and bonded to a piezoelectric actuator made out of lead zirconate titanate with a central aperture of \SI{3.1}{mm} diameter for optical access and a resonance frequency of $\SI{530}{kHz}$. The piezoelectric actuator has dimensions of $\SI{7}{mm}\times\SI{7}{mm}\times\SI{2}{mm}$ and provides a movement range of $\sim$\SI{200}{nm} along the optical axis ($\Delta z$) for fine tuning of the cavity length with picometre precision. A cryo-compatible epoxy, chosen for its high stiffness, is used to glue the actuator to a custom titanium tower. The tower geometry is designed such that its lowest mechanical resonance frequency lies above 10 kHz, far above the dominant noise band of pulse-tube cryocoolers~\cite{vadiaOpenCavityClosedCycleCryostat2021,chijiokeVibrationSpectrumPulsetube2010}. The tower is rigidly connected to a custom baseplate using titanium screws and indium washers. Tightening the titanium screws plastically deforms the indium washers, allowing the top mirror to be angularly aligned relative to the bottom mirror. This alignment is performed at room temperature outside the cryostat and is well maintained during cooldown. Titanium is used for all mechanical components close to the centre of the magnetic field. This is done to ensure compatibility with high magnetic fields and to minimise misalignment due to thermal contraction during cooldown.

A list of selected components that were used to build the described system is provided in Supplementary Note~3.

%%%%%%%%%%%%%
% Subsection 3
%%%%%%%%%%%%%
\subsection*{System diagnostics}
\label{sec:label3}

\begin{figure}
\centering
\includegraphics[width=\singlecolwidth]{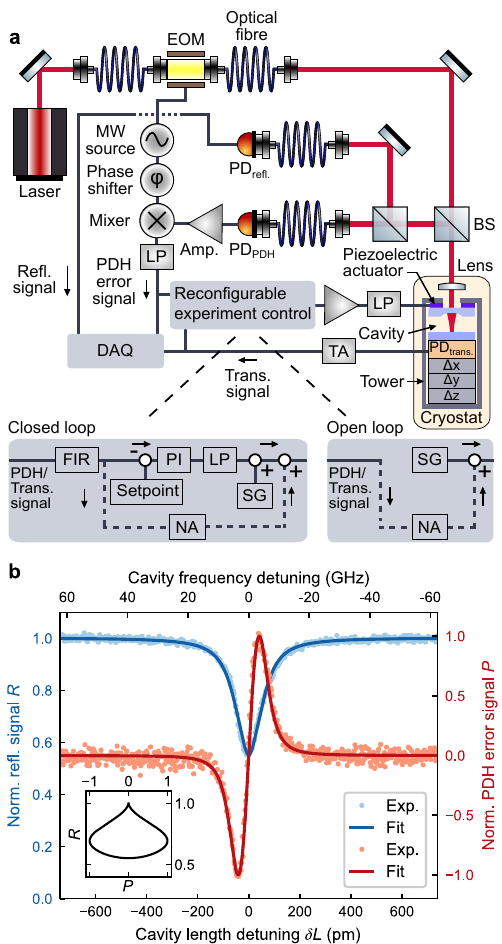}
\caption{\textbf{Optical setup and calibration.} \textbf{a}, Optical and electronic setup for characterisation of the microcavity stability and for active feedback. This system enables simultaneous detection of a transmission ($T$), reflection ($R$), and Pound-Drever-Hall (\capshort{pdh}) error signal $(P)$. Abbreviations: EOM, electro-optic modulator; PD, photodetector; TA, transimpedance amplifier; MW, microwave; BS, beam splitter; DAQ, data acquisition module; Amp., amplifier; LP, low-pass filter; FIR, finite impulse response filter; PI, proportional-integral controller; SG, signal generator; NA, network analyser; trans., transmission; refl., reflection. 
\textbf{b}, Example of simultaneously recorded signals $R$ and $P$ while the cavity resonance was swept across the laser frequency using the piezoelectric actuator. Both measured signals are calibrated against cavity length detuning $\delta L$ (see Supplementary Note~2) and fitted with a theoretical model (solid curves, see Supplementary Note~4). The inset shows a parametric plot of the fit of $R$ against the fit of $P$.}
\label{fig:3}
\end{figure}

To characterise the transfer function and cavity-length stability of the cryogenic microcavity system, we use the optical and electronic setup shown in Fig.~\ref{fig:3}a. This setup enables simultaneous measurement of the cavity transmission signal ($T)$, the reflection signal ($R$), and the \gls{pdh} signal ($P$)~\cite{black_introduction_2001,fontanaMechanicallyStableTunable2021}. Data acquisition, digital signal synthesis, and low-latency proportional-integral controller (PI) feedback are implemented using a hybrid \gls{fpga} and microprocessor board (RedPitaya StemLab 125-14), accessed through the open-source PyRPL framework~\cite{neuhausPythonRedPitaya2024}. All electrical drive signals applied to the piezoelectric actuator are amplified with a high-voltage amplifier and, when required, filtered using an analogue first-order low-pass filter with a cut-off frequency of $\SI{1.4}{kHz}$.

For stability measurements and active stabilisation, our optical probe and reference is a \SI{633}{\mathhyphen nm} external-cavity diode laser with a linewidth of $\leq\SI{300}{kHz}$, which is phase-modulated at $f_{\rm mod}=\SI{4}{GHz}$ with an electro-optic modulator before being coupled to the microcavity. The reflected light is split into two detection channels: a low-bandwidth (\SIrange{0}{200}{kHz}) photodiode records the reflected intensity, while a high-bandwidth (\SI{20}{kHz} to \SI{10}{GHz}) photodiode and subsequent electronics generate the \gls{pdh} signal (see Methods). The latter provides a linear error signal for small cavity-length displacements around resonance and can thus be used in a feedback loop for active cavity length stabilisation~\cite{black_introduction_2001}. As shown in Fig.~\ref{fig:3}b, sweeping the cavity resonance across the laser frequency using the piezoelectric actuator yields the expected symmetric reflection dip (blue data) and antisymmetric \gls{pdh} error signal (red data). To reduce the impact of vibrations on this measurement, the sweep was carried out at a rate of \SI{70}{GHz\,ms^{-1}} (\SI{0.85}{nm\,ms^{-1}}), faster than the dominant mechanical noise sources. Both measurement traces were fitted with a theoretical model (see Supplementary Note~4), showing excellent agreement. The inset displays the fitted theoretical models as a parametric curve, demonstrating that there is a one-to-one mapping between each cavity length detuning $\delta L$ and signal pair $(P(\delta L), R(\delta L))$. This unique mapping enables unambiguous readout of cavity length detuning across the full resonance, even for mechanical displacements exceeding the cavity linewidth~\cite{fontanaMechanicallyStableTunable2021}.

\begin{figure*}[t!]
\includegraphics[width=\doublecolwidth]{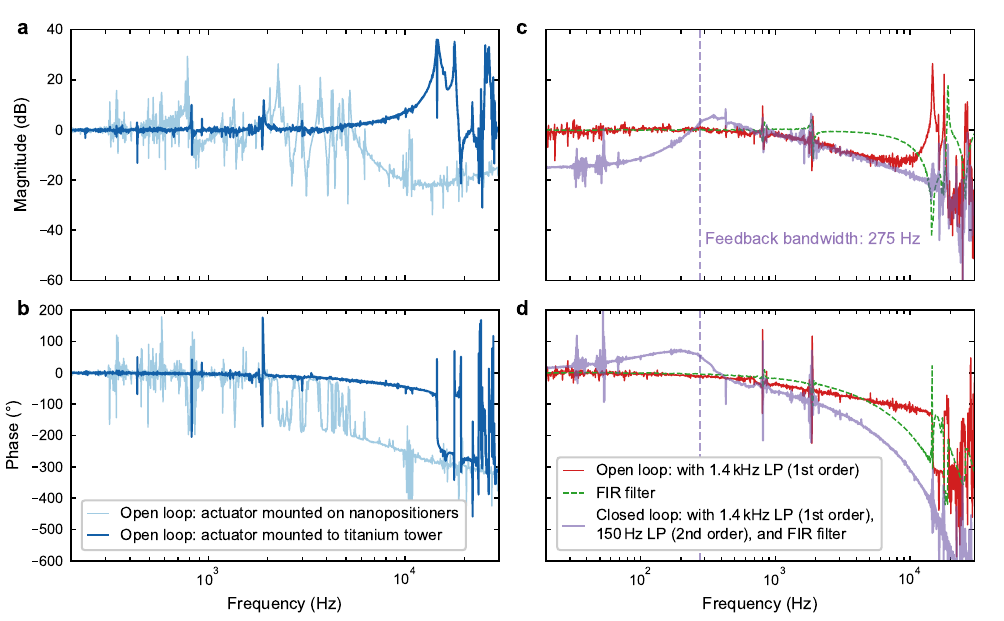}
\caption{\textbf{Transfer functions of the microcavity system at \SI[detect-all]{4}{K}.} 
The transfer functions were measured by swept-frequency sinusoidal excitation of the piezoelectric actuator. The open-loop and closed-loop measurement configurations are illustrated in Fig.~3a. \textbf{a}, \textbf{b}, Magnitude (\textbf{a}) and phase (\textbf{b}) of the open-loop transfer function with the piezoelectric actuator mounted as shown in Fig.~\ref{fig:2} (dark blue) or directly on the nanopositioner stack beneath the bottom mirror (light blue). Both configurations were measured with a \SI{532}{\mathhyphen nm} laser for which the cavity finesse is $\sim 3$.
\textbf{c}, \textbf{d}, Magnitude (\textbf{c}) and phase (\textbf{d}) of open- (red) and closed-loop (violet) transfer functions in the configuration used for active stabilisation. In both cases, a \SI{1.4}{\mathhyphen kHz} analogue low-pass (LP) filter placed after the amplifier driving the piezoelectric actuator suppresses electrical noise at the frequency of the titanium tower resonance. A second-order digital low-pass filter with a cut-off frequency of \SI{150}{Hz} within the feedback unit combined with a configurable finite impulse response (FIR) filter (green) further reduces the magnitude of the residual resonances when feedback is applied. The resulting closed-loop response shows stable noise suppression with a feedback bandwidth of \SI{275}{Hz} (see Supplementary Note~6), which also corresponds to the frequency band where residual noise dominates after passive vibration isolation. The transfer functions shown in \textbf{c} and \textbf{d} were measured with a \SI{633}{\mathhyphen nm} laser for which the cavity finesse is 3210(25).}
\label{fig:4}
\end{figure*}

Figure~\ref{fig:4} shows the cavity transfer functions, measured by swept-frequency sinusoidal excitation of the piezoelectric actuator with phase-resolved readout of the optical error signal, following the principle of a vector network analyser~\cite{neuhausPythonRedPitaya2024}. The input and output connections for the open-loop and closed-loop measurements are shown in Fig.~\ref{fig:3}a. In open loop, i.e. without active feedback, mechanical vibrations occasionally sampled the nonlinear region of the \gls{pdh} signal. For these measurements, we thus used a \SI{532}{\mathhyphen nm} laser with a linewidth of $<\SI{1}{MHz}$ outside the \gls{dbr} stopband. This intentionally broadened the cavity resonance to a spatial linewidth of $\delta L_{\mathrm{FWHM}}=\SI{82}{nm}$, well beyond the vibration amplitudes. The corresponding reflection signal guaranteed a linear displacement response for all open-loop measurements. The resulting open-loop transfer functions for two representative configurations are shown in Fig.~\ref{fig:4}a,b. The first configuration (light blue) was recorded with the piezoelectric actuator mounted beneath the bottom mirror directly on the nanopositioner stack. This measurement revealed the many resonances and antiresonances of the stacked positioners between \SI{100}{Hz} and \SI{6}{kHz}. 
The second configuration (dark blue) features the piezoelectric actuator rigidly attached to the titanium tower, as per our final design shown in Fig.~\ref{fig:2}c. This difference in mounting significantly reduced the coupling to the nanopositioner resonances while introducing coupling to only a few resonances at higher frequencies above \SI{10}{kHz}, which can be attributed to the titanium tower. To avoid control loop instabilities and to simplify controller design, we therefore used the latter mounting approach for active feedback. In general, the element controlling the cavity length should ideally be decoupled from the positioning systems, as those typically feature multiple mechanical resonances. We further inserted a first-order low-pass filter with a cut-off frequency of $\SI{1.4}{kHz}$ between the high-voltage amplifier and the piezoelectric actuator to reduce excitation of the high-frequency resonances due to electrical noise from the amplifier. The resulting open-loop transfer function is shown in Fig.~\ref{fig:4}c,d (red).\\

In closed loop, PI feedback based on the \gls{pdh} error signal reduced the cavity-length fluctuations sufficiently to ensure a linear displacement readout from the \gls{pdh} signal over the full vibration range. For these measurements, the phase-modulated \SI{633}{\mathhyphen nm} laser was thus used for greater sensitivity. Since excitation of residual resonances due to the PI controller strongly limited feedback stability, we introduced a configurable \gls{fir} filter on a second, identical \gls{fpga} board (green curve)~\cite{ryouActiveCancellationAcoustical2017} to suppress the dominant resonance–antiresonance pairs. This FIR filter was placed upstream of the PI controller and used in addition to the $\SI{1.4}{kHz}$ low-pass filter located after the high-voltage amplifier. We fitted a minimum-phase pole-zero model to the dominant resonance–antiresonance pairs of the open-loop transfer function and implemented the inverse of this fit as the \gls{fir} filter's transfer function, ensuring a causal and stable digital approximation~\cite{ryouActiveCancellationAcoustical2017}. The \gls{fir} filter output was used as the input for the network analyser, and the resulting closed-loop transfer function measurement is shown in Fig.~\ref{fig:4}c,d. The FIR filter strongly suppressed the resonances above \SI{10}{kHz}, but was less effective at compensating the residual coupling to the narrower nanopositioner resonances below \SI{2}{kHz}. The resonance near \SI{800}{Hz} was partially reduced by approximately \SI{5}{dB}, whereas the feature near \SI{2}{kHz} remained essentially unchanged. The presence of these two remaining lower-frequency resonances therefore required the use of a second-order \SI{150}{Hz} digital low-pass filter to maintain stable feedback. This filter was implemented on the \gls{fpga} on which the PI controller was running. The \SI{5}{dB} reduction of the \SI{800}{Hz} resonance nevertheless allowed the controller gain to be increased by a similar amount, resulting in a feedback bandwidth of \SI{275}{Hz} (see Supplementary Note~6). A detailed view of the FIR compensation in both frequency ranges, from \SI{700}{Hz} to \SI{2.2}{kHz} and from \SI{10}{kHz} to \SI{30}{kHz}, is provided in Supplementary Note~5.

\begin{figure*}[t!]
\includegraphics[width=\doublecolwidth]{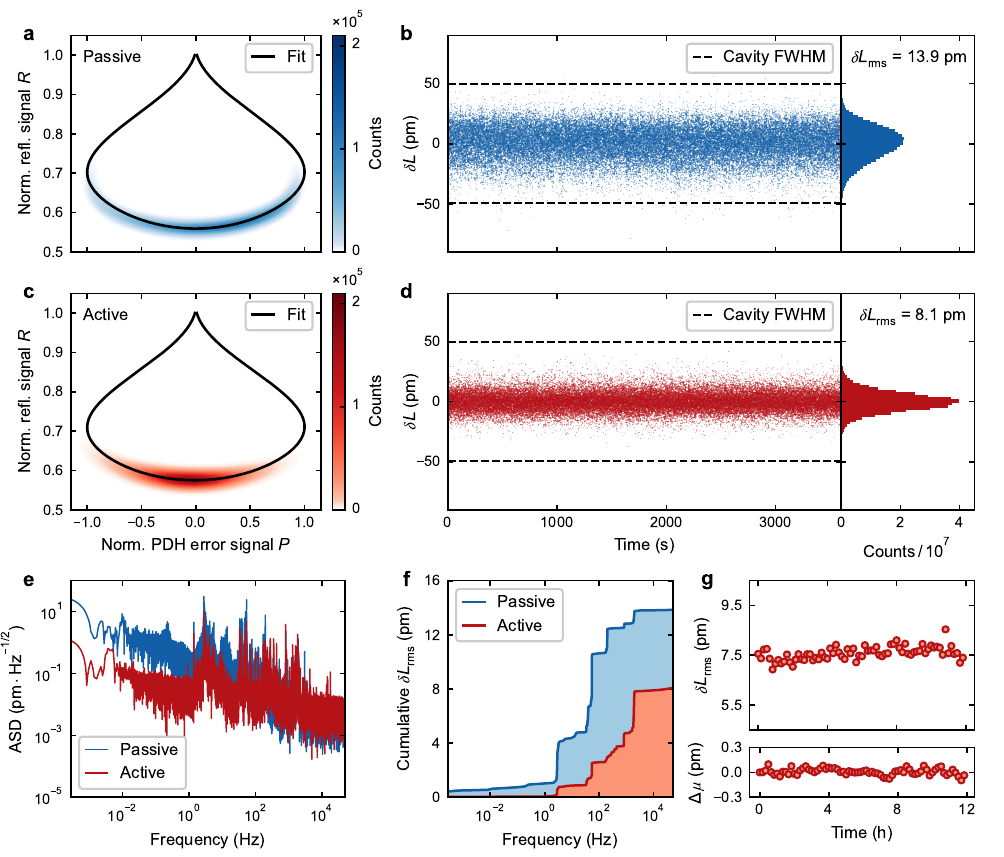}
\caption{\textbf{Performance of the microcavity system at \SI[detect-all]{4}{K}.} 
\textbf{a, c}, Parametric density maps of the cavity reflection ($R$) versus Pound-Drever-Hall (\capshort{pdh}) error signal ($P$), recorded with a sampling rate of \SI{100}{kHz} for $1\,\text{hour}$. The plot shows the joint signal distribution in $(P,R)$ space, where each 2D bin reports the number of samples falling in that region. Panel \textbf{a} shows the result with passive isolation alone, whereas panel \textbf{c} includes active stabilisation. The solid black line corresponds to the same fit that was previously shown in Fig.~\ref{fig:3}b.
\textbf{b, d}, Time traces of cavity length detuning $\delta L(t)$ converted from $(P(t), R(t))$ using the same dataset as in \textbf{a} and \textbf{c}, respectively. The accompanying histograms show the total displacement distribution over a 1-hour measurement duration. The measured root mean square (RMS) cavity length displacements are $\delta L_{\mathrm{rms}} = \SI{13.9}{pm}$ (passive, \textbf{b}) and \SI{8.1}{pm} (active, \textbf{d}).
\textbf{e}, Noise amplitude spectral densities (ASD) for the passive (blue) and active (red) configurations, computed from the time-trace data in (a) and (c), respectively.
\textbf{f}, Cumulative RMS cavity length displacement versus frequency for the passive (blue) and active (red) configurations, confirming suppression of low-frequency noise components below \SI{275}{Hz} within the active feedback bandwidth, consistent with the closed-loop response in Fig.~\ref{fig:4}c.
\textbf{g}, Interval RMS cavity length displacement in the actively-stabilised configuration measured over 12 hours in 10-s integration windows, reported every $10\,\text{minutes}$. The mean deviation ($\Delta \mu$) remains within \SI{0.1}{pm} of the lock point ($\mu$), and RMS displacements are stably maintained at \SIrange{6.9}{8.6}{pm}, demonstrating continuous long-term feedback performance.
The data for \textbf{b}, \textbf{d} were downsampled by a factor of 8000 for improved visualisation while preserving the original signal distribution. Short-term time traces of the same dataset at the full sampling rate are shown in Supplementary Note~7.
}
\label{fig:5}
\end{figure*}

\subsection*{Stability measurements}
\label{Performance}
We characterised the length stability of the microcavity at \SI{4}{K} in two situations: firstly, with passive isolation alone and, secondly, with active stabilisation engaged. The cavity reflection signal $R(t)$ and \gls{pdh} error signal $P(t)$ were recorded for \SI{1}{h} at \SI{100}{kHz} sampling rate. Figure~\ref{fig:5}a,c shows the resulting joint signal density in $(P, R)$ parameter space, superposed on the fitted parametric model (previously shown in the inset of Fig.~\ref{fig:3}b) for the passive (a) and active (c) configurations. The cavity displacement $\delta L(t)$ was then reconstructed by mapping each measured pair $(P(t), R(t))$ to the closest point on the fitted model curve $(P(\delta L), R(\delta L))$ via minimisation of the error-weighted distance~\cite{fontanaMechanicallyStableTunable2021}. The displacement estimate is defined as~\cite{fontanaMechanicallyStableTunable2021}

\begin{equation}
\delta L(t) = \underset{\delta L}{\arg \min } {\sqrt{\frac{[P(t)-P(\delta L)]^2}{\sigma_{P}^2}+\frac{[R(t)-R(\delta L)]^2}{\sigma_{R}^2}}},
\end{equation}
where $\sigma_{R}$ and $\sigma_{P}$ are the standard deviations of $R$ and $P$, measured under off-resonant conditions. 

Figure~\ref{fig:5}b,d displays the resulting time traces of the cavity length displacement $\delta L(t)$ along with histograms of $\delta L(t)$ for the passive (b) and active (d) configurations. Shorter time traces from the same dataset are presented in Supplementary Note~7. No postselection for a quiet cycle was applied in any of our analysis, enabled by strong suppression of the modulation of cavity length detunings at the pulse-tube cycle frequency (see Supplementary Note~1). Our results thus demonstrate operation at a full experimental duty cycle. The measured RMS cavity length displacement noise was $\delta L_{\rm rms}=\SI{13.9}{pm}$ with passive isolation only, reducing to \SI{8.1}{pm} with active stabilisation. This confirms that our design achieves effective mechanical decoupling from pulse-tube–induced vibrations.

Figures~\ref{fig:5}e,f show the noise amplitude spectral density (ASD) (e) and cumulative RMS noise (f) for both passive and active configurations. The dominant contributions arise from the nanopositioner stack resonances in the \SI{100}{Hz} to \SI{6}{kHz} frequency range, consistent with the open-loop transfer function measurements shown in Fig.~\ref{fig:4}. The passive isolation stage effectively attenuates vibrations across the full pulse-tube noise band (see Supplementary Note~1), with the dominant residual noise being below \SI{200}{Hz}. The active feedback suppresses noise below \SI{275}{Hz}, consistent with the feedback bandwidth determined from the closed-loop transfer function measurement (see Supplementary Note~6). The combined isolation strategy therefore provides noise reduction across the full noise spectrum of the pulse-tube cryocooler.

Lastly, in Fig.~\ref{fig:5}g we report a long-term stability measurement spanning $12\,\text{hours}$ with continuous active stabilisation. The cavity length detuning was measured every $10\,\text{minutes}$ with an acquisition time of \SI{10}{s}. Cavity length stability was maintained throughout the 12-hour experiment, with $\delta L_{\rm rms}$ ranging from \SIrange{6.9}{8.6}{pm} and a mean offset of $|\Delta \mu|<\SI{0.1}{pm}$ from the lock point $\mu$ for all measurements.

\subsection*{Stability at 2.3 K with applied magnetic field and increased feedback bandwidth}

The measurements conducted thus far had revealed residual coupling of the piezoelectric actuator to the nanopositioner resonances as the main limitation to the feedback bandwidth. We identified electrical wires from the nanopositioner stack contacting the titanium tower (Fig.~\ref{fig:2}c) as the cause for this coupling and eliminated this contact by rerouting the cables. This significantly enhanced the feedback bandwidth as will be shown in this section.

To additionally showcase the suitability of our system for magneto-optical experiments, we conducted these measurements while applying a vector magnetic field of $B=\SI{283}{mT}$ with projections along the $x$- and $z$-axis of $B_x=\SI{200}{mT}$ and $B_z=\SI{200}{mT}$, respectively. Moreover, the temperature was lowered to \SI{2.3}{K}. We used the same bottom mirror as before and a top mirror with a higher transmission of \SI{2080}{ppm} at \SI{620}{nm} (DBR coating by Evaporated Coatings, Inc.). The higher mirror transmission was chosen to broaden the cavity linewidth, such that, together with active stabilisation, the residual cavity-length fluctuations remained confined to a single side of the cavity resonance. This configuration allowed the cavity transmission $T$ to serve directly as the error signal for active stabilisation and as an unambiguous measure of cavity-length displacement $\delta L$. Compared with a \gls{pdh} readout, this \gls{sof} approach substantially simplifies the experimental implementation, as it does not require additional phase modulation and demodulation. Unlike the \gls{pdh} error signal at resonance, however, the \gls{sof} transmission signal is sensitive to fluctuations in laser power, which is why the laser power was stabilised. The cavity length was $L=\SI{4.578(2)}{\micro m}$ and the measured finesse was $F=1810(11)$, which corresponds to a spatial linewidth of $\delta L_{\mathrm{FWHM}}=\SI{171(1)}{pm}$ at \SI{620}{nm} (see Supplementary Note~8).

\begin{figure*}[t!]
\includegraphics[width=\doublecolwidth]{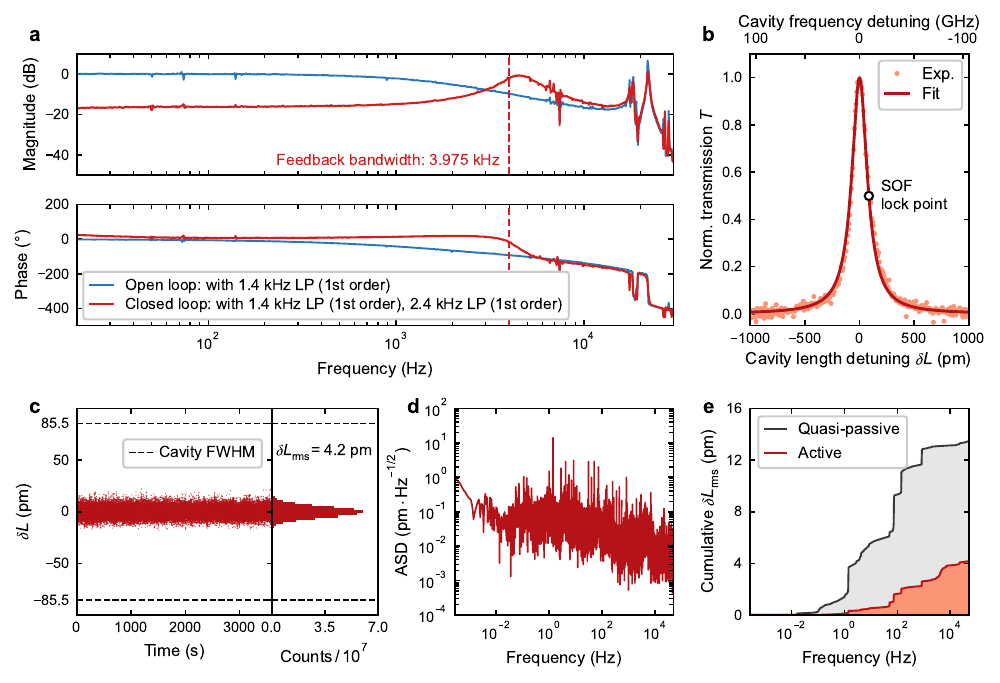}
\caption{\textbf{Performance of the tunable microcavity system at $\mathbf{2.3\,K}$ with an applied vector magnetic field.}
Measurements were performed at a magnetic field strength of $B=\SI{283}{mT}$ with $B_x=B_z=\SI{200}{mT}$.
\textbf{a,} Magnitude and phase of the open-loop (blue) and closed-loop (red) transfer functions measured using the cavity transmission signal $T$ at \SI{620}{nm} by swept-frequency sinusoidal excitation of the piezoelectric actuator. The open- and closed-loop measurement configurations are illustrated in Fig.~3a. Both configurations include a \SI{1.4}{kHz} first-order LP filter, while the closed-loop configuration additionally uses a \SI{2.4}{kHz} first-order digital LP filter. The FIR filter was not used for this measurement. The resulting feedback bandwidth is \SI{3.975}{kHz} (see Supplementary Note~6).
\textbf{b,} Normalised cavity transmission $T$, measured while sweeping the cavity resonance across the laser frequency using the piezoelectric actuator. The measured signal is fitted with a Lorentzian and calibrated in terms of cavity length detuning $\delta L$ (see Supplementary Note~8). The side-of-fringe (SOF) lock point is indicated on the positive cavity-length-detuning branch.
\textbf{c,} Cavity length displacement $\delta L(t)$ relative to the SOF lock point during $1\,\text{hour}$ of active stabilisation, together with the corresponding displacement histogram. The measured RMS cavity length detuning is $\delta L_{\mathrm{rms}}=\SI{4.2}{pm}$, without postselection.
\textbf{d, e,} Noise amplitude spectral density (ASD) for active stabilisation at \SI{2.3}{K} (\textbf{d}) and cumulative RMS cavity length displacement $\delta L_{\mathrm{rms}}$ (\textbf{e}) for active stabilisation (red) and the quasi-passive configuration (grey). In the quasi-passive configuration, a lock with a low bandwidth of \SI{2.8}{Hz} is used to compensate for slow periodic drifts introduced by the applied magnetic field and provides negligible suppression above its feedback bandwidth (see Supplementary Note~9). The high-bandwidth active feedback reduces the integrated RMS displacement from \SI{13.2}{pm} to \SI{4.2}{pm}.}

\label{fig:6}
\end{figure*}

In marked contrast to the previous open-loop transfer function in Fig.~\ref{fig:4}a, the open-loop transfer function in Fig.~\ref{fig:6}a (blue) shows no resonances or antiresonances below \SI{10}{kHz}. This increased the achievable feedback bandwidth from \SI{275}{Hz} to \SI{3.975}{kHz} (see Supplementary Note~6), without requiring the FIR filter. The \SI{150}{Hz} second-order digital LP filter used previously was replaced with a \SI{2.4}{kHz} first-order digital LP filter. The corresponding closed-loop response is shown in Fig.~\ref{fig:6}a (red).

To measure the transmission signal $T$ and identify the \gls{sof} lock point, shown in Fig.~\ref{fig:6}b, we used a low laser power of \SI{900}{nW}, measured before the cryostat window, to minimise power-induced cavity resonance shifts and associated self-locking effects (this effect was not noticeable with the previous cavity top mirror at the higher temperature of \SI{4}{K}). The cavity transmission signal was recorded while the cavity was swept across the resonance at a rate of \SI{3.4}{nm/ms}, corresponding to \SI{360}{GHz/ms}. The transmission signal was converted to cavity-length detuning $\delta L$ by numerically inverting the fitted Lorentzian on the side of the transmission peak corresponding to positive $\delta L$. 

The cavity was subsequently stabilised on this side of the resonance to the \gls{sof} lock point, and the transmission signal was recorded for $1\,\text{hour}$ at a sampling rate of \SI{100}{kHz}, see Fig.~\ref{fig:6}c. We obtain a cavity stability of $\delta L_{\mathrm{rms}}=\SI{4.2}{pm}$ without any postselection. The corresponding noise amplitude spectral density (ASD) and cumulative RMS noise are shown in Fig.~\ref{fig:6}d and e, respectively. 

Without active stabilisation, we observed slow periodic drifts of the cavity length on the timescale of several seconds while the magnetic field was applied. These slow drifts prevented a purely passive reference measurement, as the cavity did not remain on one side of the resonance in this configuration. We therefore performed a quasi-passive stability measurement using weak feedback with a bandwidth of only \SI{2.8}{Hz} (see Supplementary Note~9), sufficient to compensate these slow drifts while providing negligible suppression of vibration noise at higher frequencies. The resulting quasi-passive integrated RMS cavity-length displacement is shown in Fig.~\ref{fig:6}e and reaches a maximum of \SI{13.2}{pm}.

These measurements demonstrate that the high-bandwidth active feedback strongly suppresses the residual vibration noise transmitted through the passive vibration-isolation stages across their full frequency range, reducing the RMS cavity-length displacement from \SI{13.2}{pm} to \SI{4.2}{pm}.

%%%%%%%%%%%%%
% Discussion
%%%%%%%%%%%%%
\section*{Discussion}
\label{sec:label4}

In this work, we demonstrated picometre-level cavity-length stability inside a closed-cycle cryostat at $\SI{4}{K}$ and below, using a three-axis tunable microcavity as a vibration sensor along the $z$-axis. At \SI{4}{K}, passive isolation alone was sufficient to achieve a cavity-length displacement noise of just $\SI{13.9}{pm}$ RMS, integrated from $\SI{280}{\micro\hertz}$ to $\SI{50}{kHz}$, corresponding to a measurement time of $1\,\text{hour}$. Adding \gls{pdh} feedback further reduced this noise to $\SI{8.1}{pm}$ RMS over the same frequency band. We maintained this active feedback over $12\,\text{hours}$, with the noise staying below $\SI{8.6}{pm}$ RMS and a mean offset from the lock point of less than $\SI{0.1}{pm}$. After eliminating the mechanical coupling of the feedback element to the nanopositioner resonances, we achieved $\SI{4.2}{pm}$ RMS at \SI{2.3}{K} while applying a vector magnetic field of $B=\SI{283}{mT}$, with $B_x=B_z=\SI{200}{mT}$. These measurements demonstrate that single-axis picometre-level stability can be maintained continuously, without postselection, in a mechanically tunable cryogenic platform with multiple degrees of freedom and applied magnetic fields.

The passive isolation is provided by mechanically decoupling the cavity from the pulse-tube cryocooler via a chain of magnetic-field-compatible vibration filters. Crucially, unlike earlier reports that observed strong modulation of vibration levels at the pulse-tube cryocooler cycle frequency~\cite{fontanaMechanicallyStableTunable2021, vadiaOpenCavityClosedCycleCryostat2021}, we achieved uninterrupted, long-term operational stability without postselection. We reached this stability with unmodified commercial nanopositioners, despite their multiple mechanical resonances. This is in contrast to previous tunable cryogenic platforms, which rely on custom resonance-free positioners and less effective passive isolation schemes, rendering those designs difficult to transfer to different geometries or constrained cryogenic sample spaces~\cite{fisicaroActiveStabilizationOpenaccess2024,pallmannHighlyStableFully2023}. Future improvements in passive isolation could be realised by introducing additional rubber isolators and spring interconnects with distinct cut-off frequencies, and by damping the eigenmodes of these suspension elements, for example using copper braids.

The measurements in which the cavity length was actively stabilised show that residual mechanical coupling of the feedback element to internal resonances of the experiment module is the main limitation to the achievable feedback bandwidth. Noise and feedback signals present at these residual resonances are amplified, causing the system to ring if the feedback bandwidth or gain is set too high. In the initial microcavity implementation tested at \SI{4}{K}, the feedback actuator was mechanically decoupled from the nanopositioner stack by mounting it to a rigid titanium tower. The resonances of this tower could be compensated using a programmable FIR filter on an FPGA. The remaining coupling to the nanopositioner resonances, which were too narrow in frequency to be cancelled with the finite-length FIR filter used in this work, limited the feedback bandwidth to \SI{275}{Hz}. Rerouting wires in the microcavity module, that coupled the tower and the nanopositioner stack, removed the resonances and antiresonances below \SI{10}{kHz} and increased the achievable feedback bandwidth to \SI{3.975}{kHz}, without requiring FIR filter compensation. Further improvements should therefore focus primarily on reducing the number and quality factors of mechanical resonances and on minimising mechanical coupling between the feedback actuator and the tunable experiment module. Lower-noise electronics and an FPGA able to accommodate FIR filters with more taps would allow the residual resonance coupling to be compensated more effectively and could provide additional gains where resonances cannot be eliminated via the mechanical design. 

Our platform offers two further advantages: First, thermalisation to \SI{4}{K} and below is provided by a helium exchange gas rather than by a cold plate, making it independent of the geometry of the tunable experiment module and avoiding the experiment-specific thermal contacts that cold-plate-based designs require. Such contacts often limit the achievable thermalisation and introduce undesired mechanical coupling~\cite{herrmannLowtemperatureTunableMicrocavity2024,herrmannCoherentCouplingDiamond2024}. Second, our platform features stable free-space optical access, providing substantial freedom in the choice of beam paths and optical measurement schemes. 

Our approach is therefore readily transferable to a range of demanding tunable experiments in closed-cycle cryogenic environments. The most direct application is \gls{cqed} with solid-state emitters~\cite{janitzCavityQuantumElectrodynamics2020a,rufQuantumNetworksBased2021a}. There, a sample containing quantum emitters can be placed on the bottom mirror, whose three positioning axes can then be used to align a chosen emitter with the cavity mode. \gls{laspm}~\cite{hillenbrandVisibletoTHzNearfieldNanoscopy2025} and \gls{com}~\cite{aspelmeyerCavityOptomechanics2014} experiments often require stability on the order of tens of picometres along all three axes. Here, we demonstrated this along the $z$-axis only, but the microcavity can equally be configured to sense noise along the $x$- or $y$-axis. Adapting the isolation system to suppress that noise to a comparable level is likewise possible, facilitated by the fact that all filters except the custom spring interconnect operate at room temperature. Beyond these applications, our platform could also be extended to a wide variety of nanoscale scanning-probe microscopy approaches~\cite{birkbeckQuantumTwistingMicroscopy2025}.

Some techniques, \gls{com} experiments in particular, impose stringent vacuum requirements that are incompatible with the use of a helium exchange gas. In the present design the exchange gas provides the thermal coupling to the cold stages of the cryostat and therefore cannot be pumped out after cooldown, leaving a pressure of approximately $\SI{35}{mbar}$ in the sample chamber once cold. For such experiments, the experiment module could instead be housed in a vacuum cell, following the approach of Dubielzig et al.~\cite{dubielzigUltralowvibrationClosedcycleCryogenic2021}. The exchange gas then surrounds this cell, cools it and thermalises the module inside to the same $\SI{4}{K}$ and below. Such a cell can be evacuated and sealed at high vacuum outside the cryostat using a pinch-off tube, and cryopumping upon cooldown can then further reduce the pressure to ultra-high vacuum~\cite{dubielzigUltralowvibrationClosedcycleCryogenic2021}.

 The improvements to passive isolation and active feedback outlined above, together with extending the sensing and isolation to the $x$- and $y$-axes, would bring few-picometre RMS stability along all three axes within reach. Combined with free-space optical access, magnetic-field compatibility and three-axis tunability, this would establish experiment platforms based on closed-cycle cryostats as versatile tools for ultrastable surface-science experiments and quantum-technology applications.

%%%%%%%%%%%%%
% Methods
%%%%%%%%%%%%%
\section*{Methods}
\label{sec:Methods}

\subsection*{Microcavity mirror fabrication}
The top mirror was fabricated from a $500$-\si{\micro\metre} thick UV-fused silica substrate (UQG Optics)~\cite{strammaTinvacancyCentreDiamond2024} featuring a broad anti-reflection coating (\SIrange{400}{1100}{nm}) on one side. This substrate was diced into $\SI{5}{mm} \times \SI{5}{mm}$ chips. A square plinth with a height of \SI{100}{\micro\metre} and edge length of \SI{300}{\micro\metre} was machined into the uncoated side of the chip using a diamond saw. This geometry permits narrow air gaps between the mirror surfaces, ensuring that the substrate edges do not limit the minimum separation in the presence of angular misalignment. A spherical dimple with a radius of curvature of \SI{19}{\micro\metre} and a sagittal depth of \SI{1.5}{\micro\metre} was subsequently milled into the plinth using a focused ion beam. The bottom mirror consists of a planar UV-fused silica substrate (LaserOptik GmbH) with a thickness of $\SI{500}{\micro\metre}$ and a diameter of \SI{8}{mm}. Both substrates used for the stability measurements at \SI{4}{K} were coated (LaserOptik GmbH) with a high-reflectivity \gls{dbr} composed of 12 pairs of alternating high- ($\mathrm{Ta}_2\mathrm{O}_5$) and low- ($\mathrm{Si}\mathrm{O}_2$) refractive index materials with quarter-wavelength thickness~\cite{strammaTinvacancyCentreDiamond2024}. The transmission losses of the mirrors at $\SI{633}{nm}$ were measured to be 356 ppm (LaserOptik GmbH). The top mirror for the measurements at \SI{2.3}{K} was coated with 10 alternating pairs of the same materials and thicknesses with transmission losses measured to be 2080 ppm (Evaporated Coatings, Inc.).

\subsection*{Reflection signal, transmission signal, and Pound-Drever-Hall error signal generation}
A laser with a wavelength of \SI{633}{nm} and a linewidth of $\leq\SI{300}{kHz}$ (Toptica DL 100) was phase-modulated using a fibre-coupled electro-optic modulator (EOM, JENOPTIK PM635) driven at \SI{4}{GHz} by a synthesiser (Windfreaktech synthHD) before being coupled to the cavity. The cavity reflection signal was then split into two detection channels. A slow photodiode (FEMTO, OE-200-SI-FC, \SI{200}{kHz} bandwidth) detected the total reflected intensity, while a fast photodiode (Newport BB818-45A, bandwidth of $\SI{20}{kHz}$ to $\SI{10}{GHz}$) detected the fast modulation. The fast photodiode signal was amplified with three microwave amplifiers (Mini-Circuits ZX60-83LN-S+) and filtered with a microwave cavity bandpass filter centred at \SI{4}{GHz} (Mini-Circuits ZVBP-4000-S+). The filtered signal was then mixed (Mini-Circuits ZX05-14-S+) with a phase-shifted copy of the signal driving the \gls{eom}. The down-converted output was passed through a \SI{500}{kHz} low-pass filter (Thorlabs EF506) before being processed by an \gls{fpga} (Red Pitaya StemLab 125-14). The transmission signal was measured with a photodiode (Thorlabs FDS1010), mounted below the microcavity, and a transimpedance amplifier (FEMTO DLPCA-200, \SI{50}{kHz} bandwidth).

%%%%%%%%%%%%%
% Declarations
%%%%%%%%%%%%%

\section*{Data availability}
The data generated in this study have been deposited in the Apollo database [citation including DOI will be added here].

\section*{Code availability}

The code for analysing and plotting the data in the main manuscript is included in the Apollo database [citation including DOI will be added here].

\section*{Acknowledgements}
This work was supported in part by the ERC Advanced Grant PEDESTAL (884745) (W.G.R., M.A.), by the Quantum Computing and Simulation Hub (EP/T001062/1) through the Partnership Resource Fund PRF-09-I-06 (D.A.G., A.T.), and by the Integrated Quantum Networks Hub (EP/Z533208/1) (A.T., D.A.G., M.A.). D.A.G. acknowledges support from a Royal Society University Research Fellowship. 

\section*{Author Contributions}
W.G.R. and A.T. contributed equally to this work. W.G.R. and A.T. designed and built the experimental setup, performed the measurements, and analysed the data. A.M.S. designed early versions of the microcavity. A.D. and J.M.S. performed focused ion beam milling for the top mirror fabrication. W.G.R., A.T., M.A., and D.A.G. wrote the manuscript with input from all authors. D.A.G. and M.A. conceived and supervised the project.

\section*{Competing interests}
W.G.R., A.T., D.A.G. and M.A. declare a potential conflict of interest in relation to the intellectual property associated with this research and a pending patent application. The work described in the patent application shares elements with the research findings that are also the subject of the current paper. The title of the patent is "An insert for suspending an illuminable substrate within a cooling chamber of a closed-cycle cryostat". The application number is GB2600866.4. The geographical region is the United Kingdom. We have taken measures to ensure that all disclosed information complies with the policies and ethical guidelines of both the patenting body and the journal. All other authors declare no competing interests. 

%%%%%%%%%%%%%
% Bibliography
%%%%%%%%%%%%%

\clearpage
\onecolumngrid

%%%%%%%%%%%%%
% Supplementary Information
%%%%%%%%%%%%%
\setcounter{section}{0}
\setcounter{figure}{0}
\setcounter{equation}{0}

\renewcommand{\thesection}{\arabic{section}}
\renewcommand{\thefigure}{\arabic{figure}}
\renewcommand{\theequation}{\arabic{equation}}

\renewcommand{\figurename}{Supplementary Fig.}
\renewcommand{\thefigure}{\arabic{figure}}
\renewcommand{\thesection}{\arabic{section}}

\begin{center}
  {\Large\bfseries Supplementary Information\\[0.1em] for\\[0.4em]
   A dry cryogenic platform with picometre stability for nanoscale and quantum technologies}
\end{center}
\vspace{0.5em}

\newcommand{\authorblock}{%
\begin{center}
\begin{minipage}{0.75\textwidth}
\centering
{\fontsize{10.5}{11}\selectfont
\vspace{0.5em}
William G. Roth,\textsuperscript{1,\hyperlink{fn:equal}{*}}
Andreas Thurn,\textsuperscript{1,\hyperlink{fn:equal}{*},\hyperlink{fn:corr}{\textdagger}}
Alexander M. Stramma,\textsuperscript{1, \hyperlink{fn:paddr}{\textdaggerdbl}}
Amit R. Dhawan,\textsuperscript{2}
Jason M. Smith,\textsuperscript{2}
Dorian A. Gangloff,\textsuperscript{1,\hyperlink{fn:corr}{\textdagger}} and
Mete Atat\"ure\textsuperscript{1,\hyperlink{fn:corr}{\textdagger}}
}
\end{minipage}

{\vspace{0.6em}
\fontsize{9.5}{10}\selectfont\itshape{
$^{1}$Cavendish Laboratory, University of Cambridge, J. J. Thomson Avenue, Cambridge CB3 0US, UK\\
$^{2}$Department of Materials, University of Oxford, Parks Road, Oxford OX1 3PH, UK
}}
\end{center}
\vspace{-0.2em}%
}

\authorblock

\begingroup
\setlength{\footnotesep}{0.6\baselineskip}
\blfnotetext{\hypertarget{fn:equal}{\scalebox{0.7}{*}}} {These authors contributed equally to this work.}
\blfnotetext{\hypertarget{fn:corr}{\scalebox{0.7}{\textdagger}}}{Corresponding authors: at2116@cam.ac.uk; dag50@cam.ac.uk; ma424@cam.ac.uk.}
\blfnotetext{\hypertarget{fn:paddr}{\scalebox{0.7}{\textdaggerdbl}}}{\kern-0.25\baselineskip Present address: QuTech and Kavli Institute of Nanoscience, Delft University of Technology, Delft, The Netherlands.}
\endgroup

\section*{Supplementary Note 1: Microcavity stability without vibration isolation}
\label{sec:note_D}
Before implementing vibration isolation, we characterised the effect of cryocooler-induced vibrations on the microcavity stability. The measurement was performed with the same tunable microcavity system described in the main text, mounted with a standard measurement insert designed for cryostat bores with $\SI{49.7}{mm}$ diameter. As the mechanical noise exceeded the cavity resonance \gls{fwhm} at \SI{633}{nm}, we used a \SI{532}{nm} laser instead (M Squared Equinox), resulting in a low finesse ($F\approx3$) and a spatial linewidth of $\delta L_{\mathrm{FWHM}} =$ $\SI{82}{nm}$ due to high mirror transmission outside the DBR stopband. This broad cavity linewidth allowed the cavity to be stably tuned to a side-of-fringe configuration, providing a linear reference signal.

\begin{figure}[h]
\centering
\includegraphics[width=\singlecolwidth]{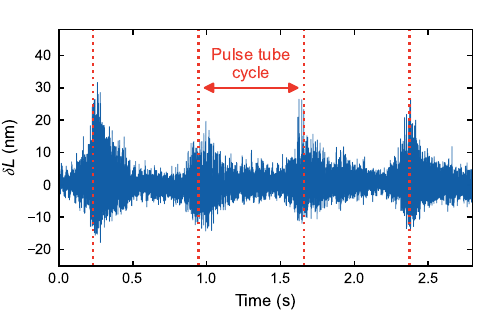}
\caption{\textbf{Time trace of cavity length detunings without vibration isolation.} Using a $\SI{532}{nm}$ laser to probe the cavity reflection signal resulted in broad reflection dips with a spatial linewidth of approximately $\delta L_{\mathrm{FWHM}}=\SI{82}{nm}$. The cavity length could therefore reliably be tuned to a side-of-fringe configuration, providing a clear reference for cavity length detunings ($\delta L$) in the presence of high-amplitude noise. The trace shows clear transients with increased noise at the start of each pulse-tube cycle.}
\label{sfig:5}
\end{figure}

Supplementary Figure~\ref{sfig:5} shows the resulting time trace, where the pulse-tube cryocooler cycle with a period of \SI{0.7}{s} is clearly visible, repeatedly exciting resonances of the microcavity system. Peak-to-peak vibration amplitudes reached \SI{45}{nm}, with insufficient ringdown before re-excitation. The corresponding vibration spectrum, shown in Supplementary Fig.~\ref{sfig:6}a, extends well into the kHz range. The resulting cumulative RMS noise is shown in Supplementary Fig.~\ref{sfig:6}b, reaching a maximum value of $\delta L_{\mathrm{rms}}=\SI{4.3}{nm}$. In comparison, after applying the vibration isolation strategy presented in this work, the vibration level was reduced by three orders of magnitude.

\begin{figure}
\centering
\includegraphics[width=\doublecolwidth]{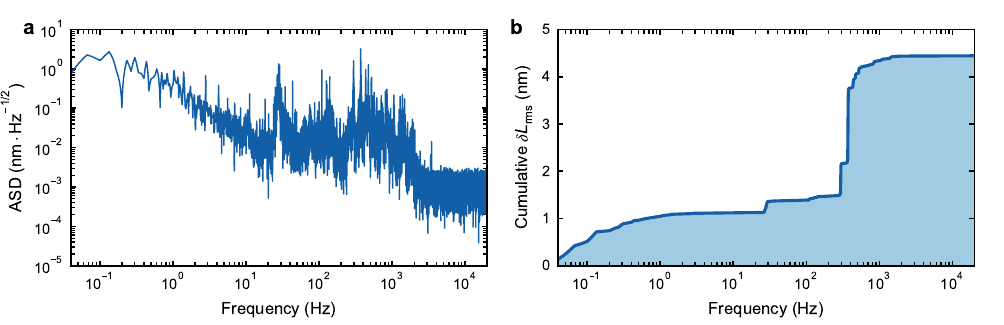}
\caption{\textbf{Spectral analysis of cavity length detunings without vibration isolation.} \textbf{a}, Amplitude spectral density (ASD) of the cavity length detunings ($\delta L$). \textbf{b}, The corresponding integrated root mean square values ($\delta L_{\mathrm{rms}}$). The dominant contributions arise from nanopositioner resonances in the range from \SI{100}{Hz} to \SI{1}{kHz}. The analysis is based on the same dataset as the time trace shown in Supplementary Fig.~\ref{sfig:5}.}
\label{sfig:6}
\end{figure}

\section*{Supplementary Note 2: Finesse and linewidth of the microcavity used for measurements at 4 K}
\label{sec:note_A}

\begin{figure}
\centering
\includegraphics[width=\singlecolwidth]{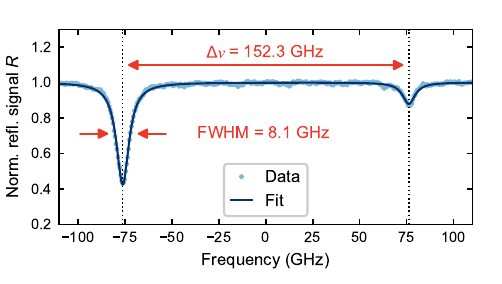}
\caption{\textbf{Measurement of the linewidth of the microcavity used for measurements at 4 K.} Two narrow-linewidth lasers at \SI{620}{nm} with a fixed relative detuning of $\Delta\nu=\SI{152.3}{GHz}$ were simultaneously coupled to the microcavity. The microcavity resonance frequency was tuned across both laser frequencies using the piezoelectric actuator while measuring the reflection signal $R$. Lorentzian fits to the reflection dips determine the microcavity linewidth, resulting in a full width at half maximum (\gls{fwhm}) of $\delta \nu=\SI{8.1(1)}{GHz}$.}
\label{sfig:1}
\end{figure}

The spatial cavity linewidth $(\delta L_{\mathrm{FWHM}})$ is given by
\begin{equation}
\delta L_{\mathrm{FWHM}} = \frac{\lambda}{2F},
\end{equation}
where $F$ is the cavity finesse and $\lambda$ is the wavelength. The spatial cavity linewidth $\delta L_{\mathrm{FWHM}}$ describes the FWHM of the reflection dip when the cavity length is scanned across the cavity resonance at a fixed laser wavelength. The cavity finesse is defined in terms of the free spectral range $\nu_{\mathrm{FSR}}$ and the spectral cavity linewidth $\delta\nu$:
\begin{equation}
F = \frac{\nu_{\mathrm{FSR}}}{\delta\nu}.
\end{equation}
Supplementary Figure~\ref{sfig:1} displays the measurement of $\delta\nu$. Two lasers (Toptica DL-SHG pro and Sacher Lasertechnik TEC 150 with AdvR SHG waveguide) with linewidths of $\leq \SI{2}{MHz}$ at \SI{620}{nm} were simultaneously coupled to the cavity using a fibre beamsplitter (Thorlabs TN632R5A2). This wavelength was chosen to utilise available narrow-linewidth sources for precise characterisation. The laser frequencies were stabilised with a wavemeter (HighFinesse WS-U) and a custom PID controller. Their frequency spacing was set to $\Delta \nu = \SI{152.3}{GHz}$ to ensure both reflection dips were clearly resolvable. The cavity resonance frequency was tuned across both laser frequencies at a rate of \SI{520}{GHz\,ms^{-1}} (corresponding to \SI{6.5}{nm\,ms^{-1}}) using the piezoelectric actuator, which is faster than the timescale of the dominant mechanical noise sources. The cavity reflection signal was recorded with a high-bandwidth photodiode (Thorlabs APD 430A, \SI{450}{MHz} bandwidth). Lorentzian fits to the reflection dips, calibrated by the known frequency spacing between their centres, yielded a spectral cavity linewidth of $\delta\nu = \SI{8.1(1)}{GHz}$.

The free spectral range $\nu_{\rm FSR}$ was obtained from cavity reflection spectra recorded under white-light illumination. A fibre-coupled LED centred at \SI{625}{nm} (Thorlabs M625F2) was coupled to the cavity, and the reflected light was analysed with a spectrometer. Supplementary Figure~\ref{sfig:2} shows the corresponding reflection dips at the same cavity length used in Supplementary Fig.~\ref{sfig:1}. Note that the shapes of the recorded spectra are modulated by the strong chromaticity of the aspheric coupling lens. From these spectra, the free spectral range was determined to be $\nu_{\mathrm{FSR}} = \SI{26.70(1)}{THz}$. This corresponds to a cavity length of $L=\SI{5.614(2)}{\micro m}$.

Combining these results, the finesse at \SI{620}{nm} is $F_{620} = 3311(24)$. Since the cavity is stabilised at \SI{633}{nm} during operation, we extrapolate these results to the locking wavelength. Accounting for the higher transmission losses due to the mirror coatings at \SI{633}{nm} compared to \SI{620}{nm}, the finesse and spatial linewidth at \SI{633}{nm} are $F_{633} = 3211(24)$ and $\delta L_{\mathrm{FWHM}} = \SI{98.5(8)}{pm}$, respectively.

\section*{Supplementary Note 3: List of components}
Table~\ref{tab:components} summarises a selection of components we used to build our system. Equivalent components with comparable properties may be substituted. Relevant physical properties are described in the main text.

\begin{table}[h]
\centering
\small
\caption{\textbf{List of components.} Selection of components that were used to build our system.}
\label{tab:components}
\renewcommand{\arraystretch}{1.2}
\begin{tabular}{@{} l l @{}}
\toprule
\textbf{Manufacturer, Part No.} & \textbf{Description} \\
\midrule
 
\multicolumn{2}{@{}l}{\textit{Passive vibration isolation}} \\
Thorlabs, PWA074 & Passive air dampers \\
RS PRO, 126-4282 & Rubber connectors \\
LewVac, EWB50KF-40 & Edge-welded bellows \\
Polycraft, K4 & Silicone \\
\midrule

\multicolumn{2}{@{}l}{\textit{Cavity assembly and positioning}} \\
Thorlabs, PA4GEH5W & Top-mirror piezo actuator\\
Thorlabs, BPA100 & High-voltage amplifier \\
Attocube, ANPx51 \& ANPz51 & Nanopositioner stack \\
Attocube, ANPx101 & Nanopositioner for aspheric lens \\
Loctite, Ablestik 451 & Cryo-compatible epoxy\\

\bottomrule
\end{tabular}
\end{table}

\section*{Supplementary Note 4: Reflection and PDH error signal model}
\label{sec:note_B}

Here, we discuss the theoretical models used in Fig.~3 to fit the \gls{pdh} error signal $P(\delta L)$ and the cavity reflection data $R(\delta L)$ as a function of cavity length detuning $\delta L$. The phase-modulated light incident on the cavity consists of a carrier at frequency $\omega$ and sidebands at $\omega\pm n\Omega$, where $\Omega$ is the modulation frequency~\cite{fontanaMechanicallyStableTunable2021} and $n\in\{ 0,1,2\}$.

The reflection signal is modelled as a composite function in which the carrier ($n=0$) and sidebands ($n=1,2$) contribute Lorentzian dips to the background reflection level. The relative depth of these dips is governed by the modulation depth of the electro-optic modulator $\beta$, scaling with the square of the Bessel functions of the first kind, $J_n(\beta)$. The reflection signal $R(\delta L)$ as a function of cavity length detuning $\delta L$ is therefore given by

\begin{equation}
\begin{split}
 R(\delta L) &= R_{\mathrm{off}} - A \bigg[ J_0^2(\beta)\mathcal{L}(\delta L; 0, \delta l) \\
 &\quad + \sum_{n=1}^{2} J_n^2(\beta)\Big( \mathcal{L}(\delta L; n\delta_s, \delta l) + \mathcal{L}(\delta L; -n\delta_s, \delta l) \Big) \bigg],
\end{split}
\label{eq:reflection_model}
\end{equation}

where $R_{\mathrm{off}}$ is the off-resonant reflection offset, $A$ is the global dip amplitude, and $\delta_s$ is the equivalent cavity length detuning corresponding to the sideband frequency spacing. The function $\mathcal{L}(\delta L; x_0, \delta l)$ is a unit-height Lorentzian centred at $x_0$, with a fixed full width at half maximum of $\delta L_{\mathrm{FWHM}}$ (see Supplementary Note 2): 

\begin{equation}
 \mathcal{L}(\delta L; x_0, \delta L_{\mathrm{FWHM}}) = \frac{1}{1+4 \left(\frac{\delta L - x_0}{\delta L_{\mathrm{FWHM}}} \right)^2}.
\end{equation}

Based on the fit to the reflection data shown in Fig.~3b, the second-order sidebands were found to be negligible. Therefore, for the modelling of the \gls{pdh} error signal, only the first-order sidebands were considered. The modelling follows Black's approach~\cite{black_introduction_2001}. Reflection off the cavity results in different complex reflection coefficients for the carrier ($R_0$) and the first-order sidebands ($R_{\pm}$). The interference signal between the carrier and the sidebands is measured with a fast detector and the resulting electrical signal is mixed with a copy of the oscillating signal that drives the phase-modulator. The mixed signal is subsequently low-pass filtered with a cut-off frequency lower than the modulation frequency. Accounting for the experimental phase delay ($\phi$) between the local oscillator and the detector signal, and a constant DC offset ($P_{\mathrm{off}}$) due to the mixing electronics, the measured \gls{pdh} error signal can be written as 
\begin{equation}
 P(\delta L) = P_{\mathrm{s}}\left[\Re(\chi)\cos(\phi)+\Im(\chi)\sin(\phi)\right]+P_{\mathrm{off}},
\end{equation}
where $P_{\mathrm{s}}$ is a scaling factor and the interference term $\chi$ is defined as
\begin{equation}
 \chi=R_0R_{+1}^*-R_0^*R_{-1}.
\end{equation}
Here, the complex reflection coefficients $R_{k}$ with $k\in\{ 0,+1,-1\}$ are calculated using
\begin{equation}
 R_k = \frac{r(e^{-i\Psi(\nu_k)}-1)}{1-r^2e^{-i\Psi(\nu_k)}},
\end{equation}
where $r$ is the amplitude reflectivity of the identical cavity mirrors, and $\Psi(\nu_k) = 2 (L+\delta L)\nu_k/c$ is the round-trip phase shift, with $c$ being the speed of light, $L$ being the cavity resonance length and $\nu_k=(\omega+k\Omega)/(2\pi)$ being the frequencies of the incident laser beam.

\begin{figure}
\centering
\includegraphics[width=\doublecolwidth]{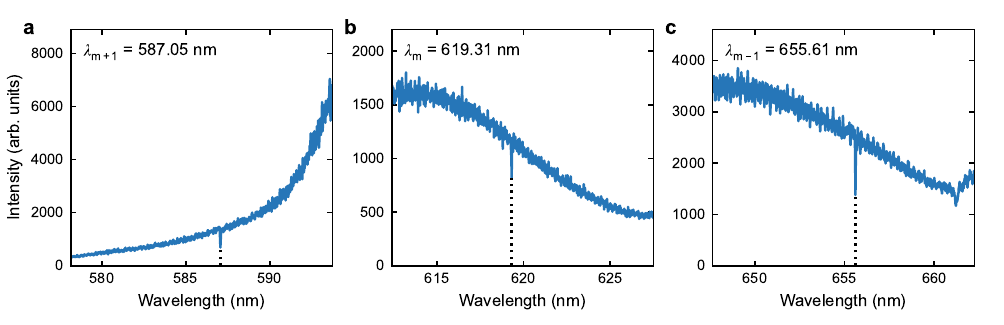}
\caption{\textbf{Measurement of the free spectral range of the microcavity used for measurements at 4 K.} \textbf{a}, \textbf{b} and \textbf{c}, Spectra of white-light reflected by the microcavity, recorded in different spectral regions to capture three neighbouring longitudinal modes. The cavity length was the same as for the measurement shown in Supplementary Fig.~\ref{sfig:1} for all three spectra. The shape of the spectra is caused by the strong chromaticity of the aspheric lens coupling to the microcavity, which results in a wavelength dependent focal shift. For the spectra \textbf{a} and \textbf{c} the position of the lens therefore had to be tuned with respect to the microcavity. The resonance wavelength corresponding to the m-th longitudinal mode is denoted by $\lambda_{\rm m}$. The three reflection dips shown in \textbf{a} to \textbf{c} occur separated by a free spectral range of $\nu_{\mathrm{FSR}}=\SI{26.7(1)}{THz}$.}
\label{sfig:2}
\end{figure}

\section*{Supplementary Note 5: Resonance compensation}
\begin{figure}
\centering
\includegraphics[width=\textwidth]{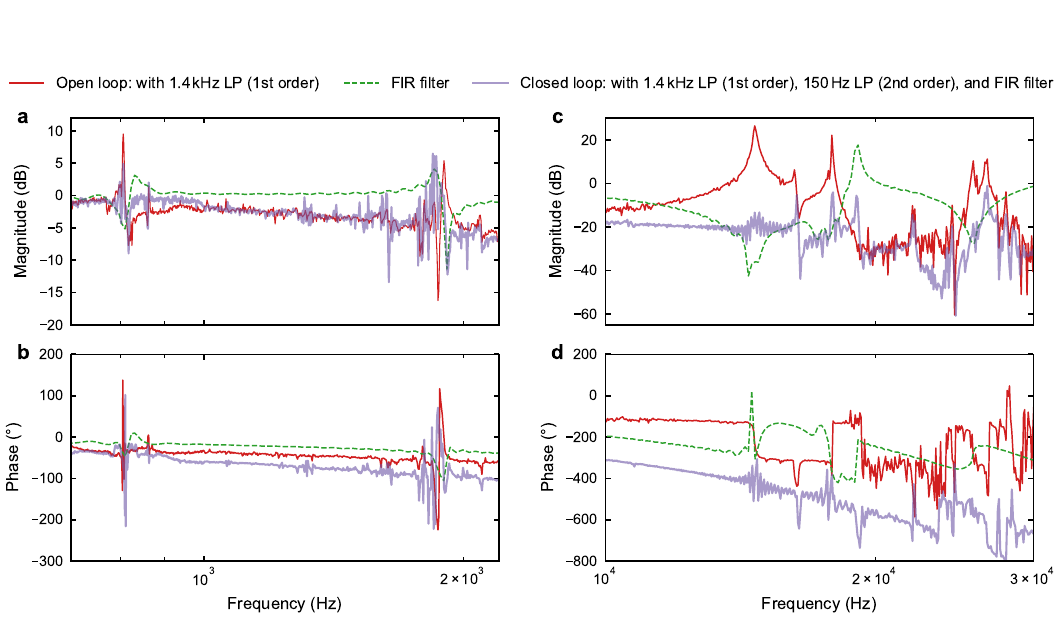}
\caption{%
\textbf{Detailed view of resonance compensation by the finite impulse response (FIR) filter.} Shown are two frequency bands of the transfer functions from Fig.~4c,d. 
\textbf{a}, 
\textbf{b}, Magnitude (\textbf{a}) and phase (\textbf{b}) of the open-loop, closed-loop and FIR filter transfer functions from \SI{700}{Hz} to \SI{2.2}{kHz}. 
\textbf{c}, 
\textbf{d}, Corresponding magnitude (\textbf{c}) and phase (\textbf{d}) from \SI{10}{kHz} to \SI{30}{kHz}.
}
\label{sfig:10}
\end{figure}

Supplementary Fig.~\ref{sfig:10} provides a detailed view of the transfer functions from Fig.~4c,d in two frequency ranges. Supplementary Fig.~\ref{sfig:10}a,b shows the magnitude and phase of the transfer functions between \SI{700}{Hz} and \SI{2.2}{kHz}, where the FIR filter only partially compensates the residual coupling to nanopositioner resonances. The resonance near \SI{800}{Hz} is reduced by approximately \SI{5}{dB}, improving the stability of the active feedback, whereas the resonance near \SI{1.96}{kHz} remains largely uncompensated. The available number of FIR coefficients is insufficient to fully reproduce such narrow spectral features, and these lower-frequency resonances therefore limit the achievable locking bandwidth. Supplementary Fig.~\ref{sfig:10}c,d shows the corresponding magnitude and phase of the transfer functions between \SI{10}{kHz} and \SI{30}{kHz}, where the broader resonances associated with the titanium tower are strongly suppressed. Although this cancellation does not increase the locking bandwidth in the present work, which remains limited by the lower-frequency resonances, it demonstrates that mechanical resonances of a specific experimental module can be effectively suppressed using a reconfigurable FIR filter.

\section*{Supplementary Note 6: Feedback bandwidth}

\begin{figure}
\centering
\includegraphics[width=\doublecolwidth]{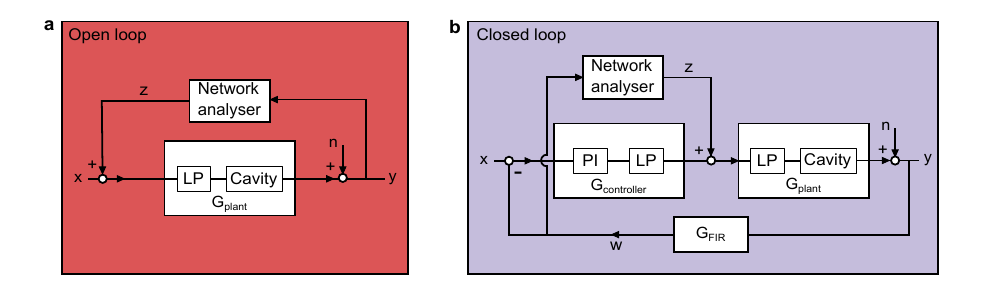}
\caption{\textbf{Block diagrams for transfer function measurements.} The block diagrams illustrate the signal flow through the system, including the network analyser. \textbf{a}, Configuration used to measure the open-loop transfer function of the plant ($G_{\mathrm{plant}}$), consisting of a \SI{1.4}{kHz} low-pass (LP) filter and the cavity. Measurement noise $n$ is added at the output to account for all noise sources up to that point in the signal chain. The network analyser adds a swept-frequency sinusoidal signal $z$ to the plant input parameter $x$ and measures its response $y$. \textbf{b}, Configuration used to measure the closed-loop transfer function $G_{\mathrm{closed\,loop}}$. The feedback loop consists of a controller, a plant, and an optional finite impulse response (FIR) filter. The controller and the FIR filter are described by transfer functions $G_{\mathrm{controller}}$ and $G_{\mathrm{FIR}}$, respectively. If the FIR filter is not used, we set $G_{\mathrm{FIR}}=1$. The controller comprises a proportional-integral (PI) stage and a configurable LP filter. The network analyser applies a swept-frequency sinusoidal signal $z$ to the input of the plant $x$ and measures the FIR-filtered response $w$.}
\label{sfig:8}
\end{figure}

Supplementary Fig.~\ref{sfig:8} shows the block diagrams used to measure the open-loop and closed-loop transfer functions presented in Fig.~4 and Fig.~6a of the main text. In the open-loop configuration, $x$ denotes a constant voltage offset, see Supplementary Fig.~\ref{sfig:8}a. This voltage offset is generated by a signal generator (SG), as indicated in Fig.~3a, and is used to bring the cavity into resonance with the laser. In the closed-loop configuration, $x$ instead denotes the setpoint of the proportional-integral (PI) controller, see Supplementary Fig.~\ref{sfig:8}b. For the measurements presented in Fig.~4c,d, the network analyser signal was measured after the finite impulse response (FIR) filter rather than directly at the plant output. This configuration was chosen to explicitly capture the contribution of the FIR filter to the measured closed-loop transfer function $G_{\mathrm{closed\,loop}}$. In contrast, the measurements presented in Fig.~6a were performed without the FIR filter. The following derivation applies to both configurations, with $G_{\mathrm{FIR}}=1$ for the measurements presented in Fig.~6a.

To determine the feedback bandwidth, we express the output signal $y$ in terms of the setpoint $x$, the signal $z$ injected by the network analyser, and the noise $n$ as
\begin{equation}
y =
\frac{G_{\mathrm{controller}}G_{\mathrm{plant}}}{1+L}x
+\frac{G_{\mathrm{plant}}}{1+L}z
+\frac{1}{1+L}n,
\label{eq:loop_output}
\end{equation}
where
\begin{equation}
L=G_{\mathrm{controller}}G_{\mathrm{plant}}G_{\mathrm{FIR}}
\end{equation}
is the loop gain. For the configuration used in Fig.~4c,d, the network analyser measures the FIR-filtered output $w=G_{\mathrm{FIR}}y$, giving
\begin{equation}
w =
\frac{L}{1+L}x
+\frac{G_{\mathrm{plant}}G_{\mathrm{FIR}}}{1+L}z
+\frac{G_{\mathrm{FIR}}}{1+L}n.
\end{equation}
For the measurements in Fig.~6a, no FIR filter was used and hence $G_{\mathrm{FIR}}=1$, such that $w=y$.
The network analyser isolates the component of $w$ that is coherent with the injected signal $z$ at frequency $f$, independently of the constant setpoint $x$ and the incoherent noise $n$. The measured closed-loop transfer function is therefore
\begin{equation}
G_{\mathrm{closed\,loop}}
=\frac{w}{z}
=\frac{G_{\mathrm{plant}}G_{\mathrm{FIR}}}{1+L}.
\end{equation}
The sensitivity function $S$, defined as
\begin{equation}
S=\frac{1}{1+L},
\end{equation}
quantifies the suppression of disturbances entering at the plant output by the feedback loop~\cite{ryouActiveCancellationAcoustical2017, bechhoeferFeedbackPhysicistsTutorial2005}, as shown in Eq.~\ref{eq:loop_output}. Comparing the expressions for $S$ and $G_{\mathrm{closed\,loop}}$ gives
\begin{equation}
S=\frac{G_{\mathrm{closed\,loop}}}
 {G_{\mathrm{plant}}G_{\mathrm{FIR}}}.
\end{equation}
Thus, the sensitivity can be determined directly from the measured transfer functions. For the measurements in Fig.~6a, this expression reduces to $S=G_{\mathrm{closed\,loop}}/G_{\mathrm{plant}}$ because $G_{\mathrm{FIR}}=1$. Finally, the corresponding loop gain is obtained from
\begin{equation}
L=\frac{1}{S}-1.
\end{equation}

\subsection{Sensitivity and loop gain calculation for the feedback loop applied at 4~K}

The resulting magnitudes of the sensitivity and loop gain corresponding to the data shown in Fig.~4c,d at 4~K are presented in Supplementary Fig.~\ref{sfig:9}. For this feedback loop, a second-order LP filter with a cut-off frequency of \SI{150}{Hz} was used. The sensitivity exhibits clear noise suppression up to \SI{240}{Hz}, followed by a servo bump extending from \SI{240}{Hz} to \SI{1}{kHz}. We define the feedback bandwidth as the unity-gain frequency ($f_{\mathrm{UGF}}$), i.e. the frequency at which $|L(f_{\mathrm{UGF}})|=\SI{0}{dB}$, yielding $f_{\mathrm{UGF}}=\SI{275}{Hz}$.

\subsection{Sensitivity and loop gain calculation for the feedback loop applied at 2.3~K}

The resulting magnitudes of the sensitivity and loop gain corresponding to the data shown in Fig.~6a at 2.3~K are presented in Supplementary Fig.~\ref{sfig:13}. For this feedback loop, a first-order LP filter with a cut-off frequency of \SI{2.4}{kHz} was used, while no FIR filter was applied. The sensitivity exhibits clear noise suppression up to \SI{3}{kHz}, followed by a servo bump extending from \SI{3}{kHz} to \SI{10}{kHz}. The unity-gain frequency is $f_{\mathrm{UGF}}=\SI{3.975}{kHz}$.

\begin{figure}
\centering
\includegraphics[width=\singlecolwidth]{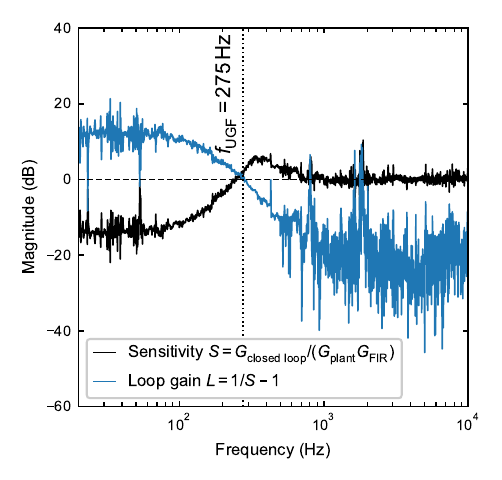}
\caption{\textbf{Magnitude of the sensitivity and loop gain of the feedback loop used for active stabilisation at 4 K.} The sensitivity $S$ and loop gain $L$ were calculated from the measured transfer functions in Fig.~4c,d as described in the text. The unity-gain frequency $f_{\mathrm{UGF}}=\SI{275}{Hz}$ is marked by a vertical dotted line. The sensitivity function exhibits a clear suppression region up to \SI{240}{Hz}, followed by a servo bump extending from \SI{240}{Hz} to \SI{1}{kHz}.}
\label{sfig:9}
\end{figure}

\begin{figure}
\centering
\includegraphics[width=\singlecolwidth]{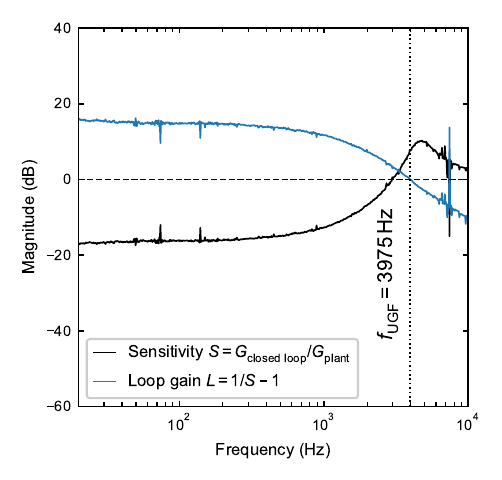}
\caption{\textbf{Magnitude of the sensitivity and loop gain of the feedback loop used for active stabilisation at 2.3 K.} The sensitivity $S$ and loop gain $L$ were calculated from the measured transfer functions in Fig.~6a as described in the text. The unity-gain frequency $f_{\mathrm{UGF}}=\SI{3.975}{kHz}$ is marked by a vertical dotted line. The sensitivity function exhibits a clear suppression region up to \SI{3}{kHz}, followed by a servo bump extending from \SI{3}{kHz} to \SI{10}{kHz}.}
\label{sfig:13}
\end{figure}

\section*{Supplementary Note 7: Detailed time traces}
\label{sec:note_C}

Figure~5 in the main text presents 1-hour time series of the cavity length detuning $\delta L$, calculated from the simultaneously recorded reflection and \gls{pdh} error signals. Due to the high sampling rate (\SI{100}{kHz}) used for this measurement, the time series data in Fig.~5 were downsampled by a factor of 8000 for improved visualisation. In this section, we therefore present time series from the same data set at the full sampling rate, but for shorter time windows. To show the underlying raw data, we include the time series of the reflection signal $R(t)$ and the \gls{pdh} error signal $P(t)$ alongside the cavity length detuning $\delta L(t)$.

Supplementary Figure~\ref{sfig:3} presents all three time series over a \SI{200}{ms} timeframe. For passive isolation, as shown in Supplementary Fig.~\ref{sfig:3}a,c,e, most residual vibrations occur at frequencies of approximately \SI{200}{Hz} or below (periods $\geq\SI{5}{ms}$). This is expected, as the efficiency of mechanical low-pass filters decreases at lower frequencies. With active stabilisation, these low-frequency components are effectively suppressed, as shown in Supplementary Fig.~\ref{sfig:3}b,d,f.

Supplementary Figure~\ref{sfig:4} displays all three time series over a \SI{5}{s} window, encompassing several pulse-tube cycles that occur with a period of \SI{0.7}{s}. Notably, for both the passive (Supplementary Fig.~\ref{sfig:4}a,c,e) and active (Supplementary Fig.~\ref{sfig:4}b,d,f) configurations the periodic modulation of mechanical noise typically associated with the pulse-tube cycle is much less pronounced compared to the standard configuration without vibration isolation (see Supplementary Note~1). This demonstrates that the microcavity is efficiently decoupled from the mechanical vibrations generated by the pulse-tube cryocooler.
\newline

\begin{figure*}
\centering
\includegraphics[width=\doublecolwidth]{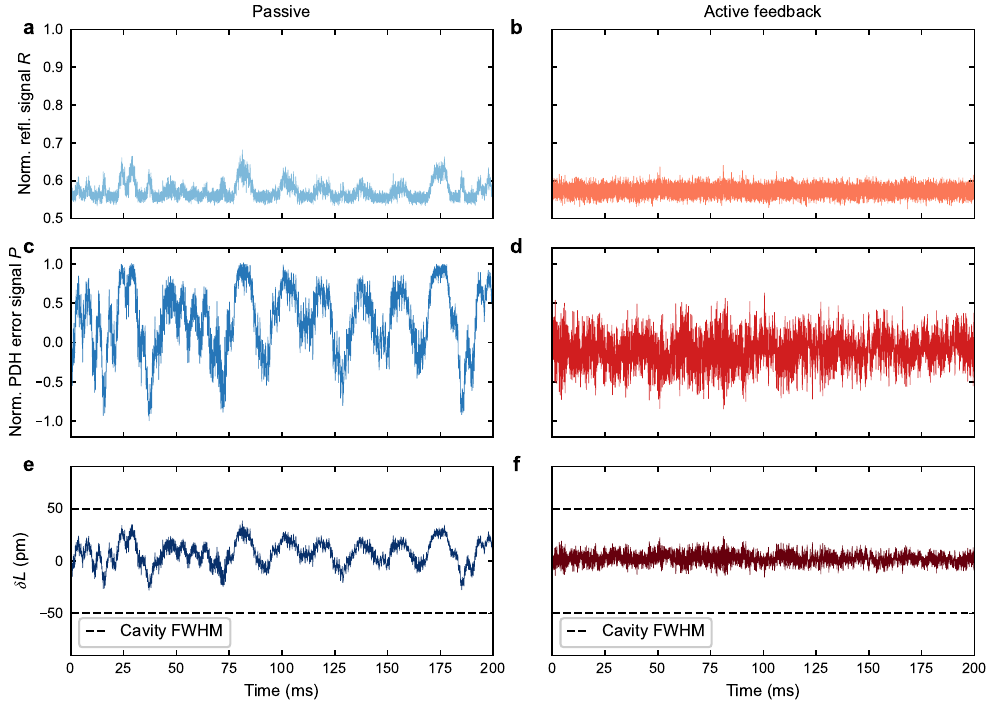}
\caption{\textbf{Sub-second time traces of microcavity-stability measurements.}
Representative \SI{200}{ms} time traces from the \SI{1}{h} dataset shown in Fig.~5, recorded with passive isolation alone (\textbf{a}, 
\textbf{c} and 
\textbf{e}) and with active feedback applied (\textbf{b}, 
\textbf{d} and 
\textbf{f}). 
\textbf{a}, 
\textbf{b}, Reflection signal $R$. 
\textbf{c}, 
\textbf{d}, Pound-Drever-Hall (\gls{pdh}) error signal $P$. 
\textbf{e}, 
\textbf{f}, Cavity-length deviation $\delta L$, obtained from $R$ and $P$ by the method described in the main text. The dashed lines indicate the full width at half maximum (\gls{fwhm}) of the cavity resonance, expressed as an equivalent cavity-length change.
The active feedback significantly attenuates residual vibrations transmitted by the passive isolation system across a bandwidth of $\SI{275}{Hz}$, see also Supplementary Note~6.}
\label{sfig:3}
\end{figure*}

\begin{figure*}
\centering
\includegraphics[width=\doublecolwidth]{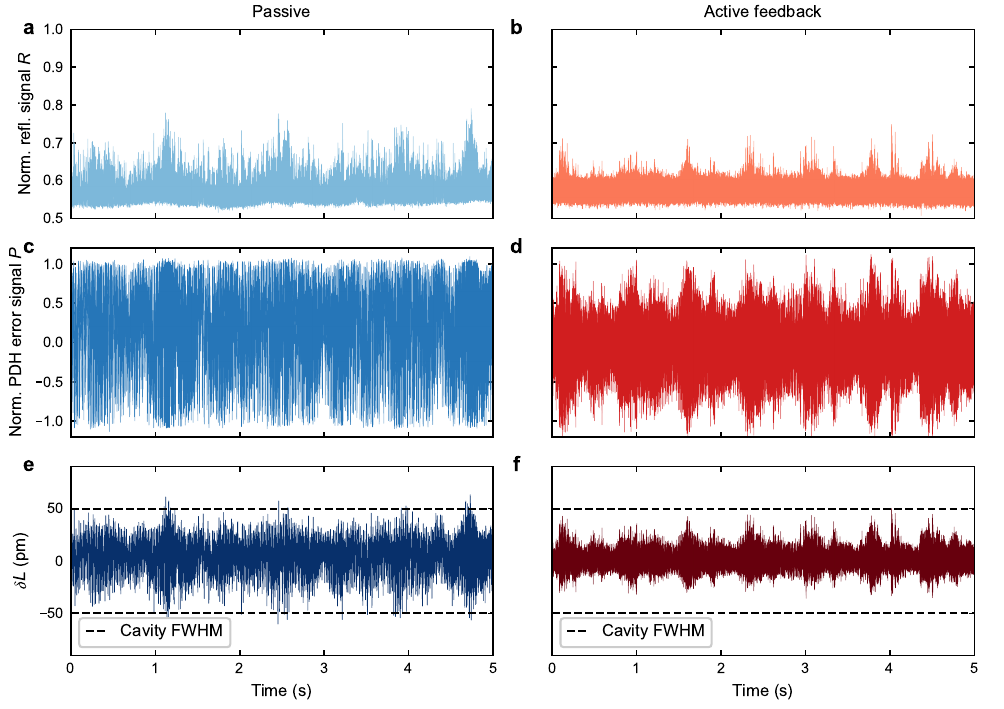}
\caption{\textbf{5-second time traces of cavity-stability measurements.} Representative \SI{5}{\mathhyphen s} time traces of the \SI{1}{h} dataset shown in Fig.~5, recorded with passive isolation alone (\textbf{a}, \textbf{c} and \textbf{e}) and with active feedback applied (\textbf{b}, \textbf{d} and \textbf{f}). \textbf{a}, \textbf{b}, Reflection signal $R$. \textbf{c}, \textbf{d}, Pound-Drever-Hall (\gls{pdh}) error signal $P$. \textbf{e}, \textbf{f}, Cavity length deviation $\delta L$, obtained from $R$ and $P$ by the method described in the main text. The dashed lines indicate the full width at half maximum (\gls{fwhm}) of the cavity resonance, expressed as an equivalent cavity-length change.}
\label{sfig:4}
\end{figure*}

\section*{Supplementary Note 8: Finesse and linewidth of the microcavity used for measurements at 2.3 K }

The calculation of the spatial cavity linewidth of the microcavity used for the measurements at 2.3 K follows the same method outlined in Supplementary Note 2. 
Supplementary Figure~\ref{sfig:11} displays the measurement of the spectral cavity linewidth $\delta\nu$. Two lasers (Toptica DL-SHG pro and MogLabs LDL 1220 with AdvR SHG waveguide) with linewidths of $\leq \SI{2}{MHz}$ at \SI{620}{nm} were simultaneously coupled to the cavity using a fibre beamsplitter (Thorlabs TN632R5A2). The laser frequencies were stabilised with a wavemeter (HighFinesse WS-U) and a custom PID controller. Their frequency spacing was set to $\Delta \nu = \SI{283.6}{GHz}$ to ensure both transmission peaks were clearly resolvable. The cavity resonance frequency was swept across both laser frequencies at a rate of \SI{278}{GHz\,ms^{-1}} (corresponding to \SI{2.5}{nm\,ms^{-1}}) using the piezoelectric actuator, which is faster than the timescale of the dominant mechanical noise sources. The optical power per laser incident on the cavity was limited to 900 nW to minimise power-induced cavity resonance shifts and associated self-locking during the measurement. The cavity transmission signal was recorded with a photodiode (Thorlabs FDS1010) mounted below the microcavity as shown in Fig.~3a of the main text and a transimpedance amplifier (FEMTO DLPCA-200, 50 kHz bandwidth). Lorentzian fits to the transmission peaks, calibrated using the known frequency spacing between their centres, yielded a spectral cavity linewidth of $\delta\nu = \SI{18.1(1)}{GHz}$.

The free spectral range $\nu_{\rm FSR}$ was obtained from cavity reflection spectra recorded under white-light illumination. A fibre-coupled LED centred at \SI{625}{nm} (Thorlabs M625F2) was coupled to the cavity, and the reflected light was analysed with a spectrometer. Supplementary Figure~\ref{sfig:12} shows the corresponding reflection dips at the same cavity length used in Supplementary Fig.~\ref{sfig:11}. Note that the shapes of the recorded spectra are modulated by the strong chromaticity of the aspheric coupling lens. From these spectra, the free spectral range was determined to be $\nu_{\mathrm{FSR}} = \SI{32.7(2)}{THz}$. This corresponds to a cavity length of $L=\SI{4.587(2)}{\micro m}$.
Combining these results, we obtain a finesse of $F_{620} = 1812(11)$ and a spatial linewidth of $ \delta L_{\mathrm{FWHM}} = \SI{171(1)}{pm}$ at \SI{620}{nm}.

\begin{figure}
\centering
\includegraphics[width=\singlecolwidth]{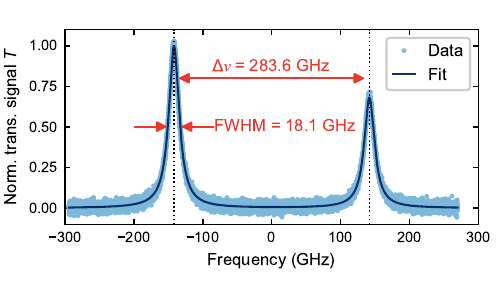}
\caption{\textbf{Measurement of the linewidth of the microcavity used for measurements at 2.3 K.} Two narrow-linewidth lasers at \SI{620}{nm} with a fixed relative detuning of $\Delta\nu=\SI{283.6}{GHz}$ were simultaneously coupled to the microcavity. The microcavity resonance frequency was tuned across both laser frequencies using the piezoelectric actuator while measuring the transmission signal $T$. Lorentzian fits to the transmission peaks determine the microcavity linewidth, resulting in a full width at half maximum (\gls{fwhm}) of $\delta \nu=\SI{18.1(1)}{GHz}$.}
\label{sfig:11}
\end{figure}

\begin{figure}
\centering
\includegraphics[width=\doublecolwidth]{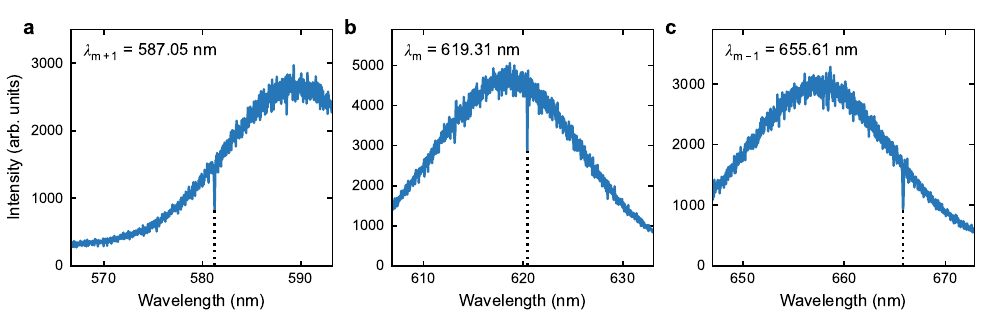}
\caption{\textbf{Measurement of the free spectral range of the microcavity used for measurements at 2.3 K.} \textbf{a}, \textbf{b} and \textbf{c}, Spectra of white-light reflected by the microcavity, recorded in different spectral regions to capture three neighbouring longitudinal modes. The cavity length was the same as for the measurement shown in Supplementary Fig.~\ref{sfig:11} for all three spectra. The shape of the spectra is caused by the strong chromaticity of the aspheric lens coupling to the microcavity, which results in a wavelength dependent focal shift. For the spectra \textbf{a} and \textbf{c} the position of the lens therefore had to be tuned with respect to the microcavity. The resonance wavelength corresponding to the m-th longitudinal mode is denoted by $\lambda_{\rm m}$. The three reflection dips shown in \textbf{a} to \textbf{c} occur separated by a free spectral range of $\nu_{\mathrm{FSR}}=\SI{32.7(2)}{THz}$.}
\label{sfig:12}
\end{figure}

\section*{Supplementary Note 9: Transfer function and feedback bandwidth of the quasi-passive configuration}

\begin{figure}
\centering
\includegraphics[width=\textwidth]{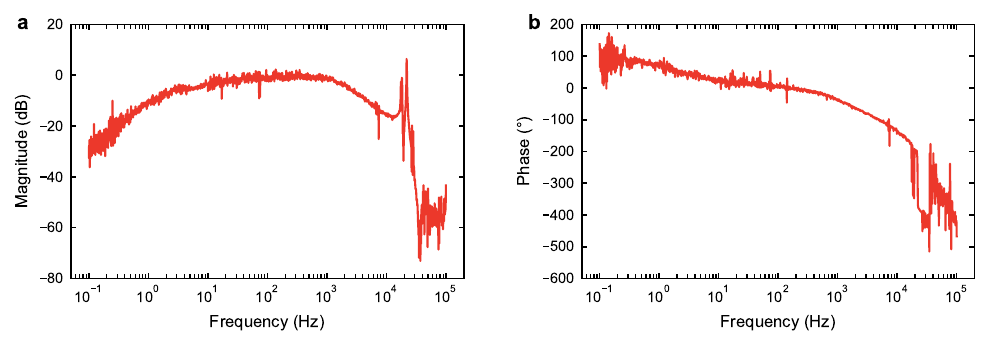}
\caption{\textbf{Closed-loop transfer function of the quasi-passive configuration.} \textbf{a, b,} Magnitude (\textbf{a}) and phase (\textbf{b}) of the measured closed-loop transfer function of the weak feedback loop used to characterise the vibration noise in the absence of high-bandwidth active feedback. The measurement was carried out at \SI{2.3}{K} with a magnetic field applied. The feedback loop shows suppression at frequencies below the bandwidth of \SI{2.8}{Hz}.}
\label{sfig:14}
\end{figure}

\begin{figure}
\centering
\includegraphics[width=\singlecolwidth]{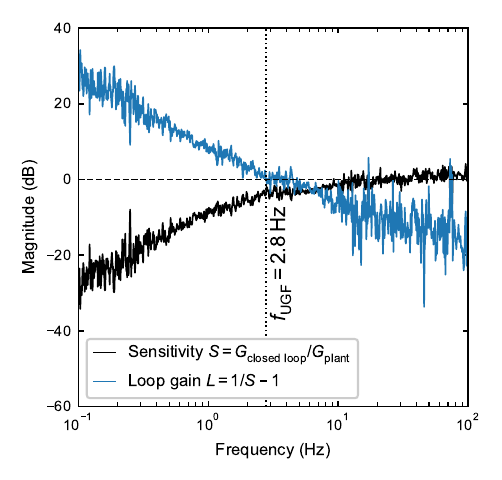}
\caption{\textbf{Sensitivity and loop gain of the quasi-passive feedback loop.} The sensitivity $S$ and loop gain $L$ were calculated from the measured closed-loop transfer function presented in Supplementary Fig.~\ref{sfig:14}, assuming a flat plant response below \SI{10}{Hz}. The unity-gain frequency, $f_{\mathrm{UGF}}=\SI{2.8}{Hz}$, is indicated by the vertical dotted line. The sensitivity shows suppression of cavity-length fluctuations below the feedback bandwidth and approaches unity above it without a pronounced servo bump.}
\label{sfig:15}
\end{figure}

Supplementary Figures~\ref{sfig:14} and \ref{sfig:15} characterise the feedback loop used in the quasi-passive configuration at \SI{2.3}{K} with applied magnetic field. The loop suppresses low-frequency drifts in cavity length below a feedback bandwidth of \SI{2.8}{Hz}, while leaving the vibration amplitudes at higher frequencies essentially unchanged. This low feedback bandwidth was realised using a \SI{4}{Hz} digital low-pass filter together with small proportional ($p$) and integral ($i$) gains implemented on the reconfigurable experiment module.

\end{document}